\documentclass[twocolumn,pra]{revtex4-2}
\usepackage{amsmath}
\usepackage{amssymb}
\usepackage{graphicx}
\usepackage{dcolumn}
\usepackage{bm}
\usepackage{float}
\usepackage{mathrsfs}
\usepackage{braket}
\usepackage{hyperref}
\usepackage{mathtools}
\usepackage{comment}
\usepackage{yfonts}
\usepackage{tabularx}

\newcolumntype{C}{>{\centering\arraybackslash}X}
\newcolumntype{L}{>{\raggedright\arraybackslash}X}
\newcolumntype{R}{>{\raggedleft\arraybackslash}X}

\mathtoolsset{showonlyrefs=false}

\hypersetup{
 colorlinks = true,
 linkcolor = {blue},
 citecolor = {blue},
 urlcolor = {blue},
}

\usepackage[whole]{bxcjkjatype}
\usepackage[normalem]{ulem}
\usepackage[usenames,dvipsnames]{xcolor}
\usepackage{soul}
\renewcommand{\d}{\dagger}

\usepackage{hyperref}

\begin{document}
\title{Lie algebraic quantum phase reduction based on heterodyne detection}
\author{Wataru Setoyama}
\email{setoyama@biom.t.u-tokyo.ac.jp}
\affiliation{
	Graduate School of Information Science and Technology,\\
	The University of Tokyo, Tokyo 113-8656, Japan
}

\author{Yoshihiko Hasegawa}
\email{hasegawa@biom.t.u-tokyo.ac.jp}
\affiliation{
 Graduate School of Information Science and Technology,\\
	The University of Tokyo, Tokyo 113-8656, Japan
}

\date{\today}

\begin{abstract}
Measurement backaction inherently alters observed dynamics in quantum physics.
In the realm of quantum synchronization, this backaction induces a phase bias, making the assessment of synchronization critically dependent on the choice of the observables.
In this study, we extend the quantum phase reduction approach~\href{https://link.aps.org/doi/10.1103/PhysRevLett.132.093602}{[PhysRevLett.132.093602]} into heterodyne detection, offering a comprehensive theoretical framework for analyzing quantum synchronization dynamics through uniform continuous measurement over all possible quadrature observables.
This method averages out the backaction, allowing for unbiased evaluation of synchronization between quantum oscillators while avoiding measurement-induced phase bias.
Furthermore, by defining the phase and limit-cycle solution independently of specific observables, our proposed method consistently adapts to the scenario where the observables are freely modified during the time evolution.
Through simulations of noise-induced synchronization, our method reveals that the number of phase clusters between oscillators is restricted by their bosonic levels.
\end{abstract}

\maketitle
\section{INTRODUCTION}
Synchronization, the alignment of an internal rhythm with external perturbations, is ubiquitous in nature, seen in firing neurons, oscillating laser beams, and Josephson junctions~\cite{Kuramoto1984, winfree2001geometry, izhikevich2007dynamical, strogatz2018nonlinear}.
With the recent advancements in quantum technology, nanoscale oscillators have become a subject of study, encompassing superconducting circuits~\cite{PhysRevLett.110.097002, PhysRevA.97.013811}, optomechanical oscillators~\cite{PhysRevX.4.011015,Weiss_2016}, atomic systems~\cite{PhysRevE.78.011108,PhysRevLett.114.103601}, and quantum thermal machines~\cite{PhysRevE.70.046110,Rezek_2006,PhysRevE.96.062120,PhysRevE.101.020201}. 
The study of synchronization in nanoscale systems offers significant promise for advancing quantum technologies such as quantum metrology, communication, cryptography, and clock~\cite{crescini2023evidence,LADD20181,doi:10.1063/5.0061478,PhysRevX.11.011046}.
For instance, investigations into quantum clocks have demonstrated that the continuously monitored coupled quantum oscillators can yield more precise temporal information than classical clocks~\cite{PhysRevX.11.011046}.
Thus, investigating synchronization within the quantum regime presents substantial technological possibilities.
Regarding this issue, the quantum models of limit-cycle oscillators (i.e., nonlinear systems exhibiting self-sustained oscillation) are proposed, including qubit and spin oscillators~\cite{PhysRevLett.121.053601, PhysRevA.101.062104} and quantum van der Pol oscillators~\cite{PhysRevLett.111.234101,PhysRevLett.112.094102}.
Additionally, experimental research has shown evidence of quantum synchronization in limit-cycle oscillators in laboratory conditions~\cite{PhysRevLett.125.013601,doi:10.1126/science.abo3382,PhysRevResearch.2.023026}.

Phase reduction is a standard theoretical scheme for studying synchronization dynamics, reducing weakly perturbed multi-dimensional limit-cycle dynamics to one-dimensional phase dynamics~\cite{Kuramoto1984,winfree2001geometry}.
Using Lie algebra, we introduced phase reduction method to quantum systems, providing a comprehensive framework for analyzing synchronization of quantum limit-cycle oscillators~\cite{PhysRevLett.132.093602}. 
In a continuous measurement scheme of their coupled environment, quantum trajectories of these oscillators come to follow a stochastic Schr\"{o}dinger equation \cite{Gisin_1992, gardiner2004quantum, barchielli2009quantum}.
Provided that the disturbance of quantum noise is weak, these trajectories fluctuate around a limit-cycle solution.
Based on the Lie algebraic framework, we calculated the phase response curves to weak perturbations represented by unitary transformations.
Thus, a quantum phase equation was derived from a stochastic Schr\"{o}dinger equation of a quantum limit-cycle oscillator.
This approach is versatile from classical to quantum regimes and applicable to general quantum limit-cycle oscillators, including qubits and spins, lacking classical limit-cycle analogues.
Note that the extension of phase reduction to quantum systems was pioneered in Ref.~\cite{PhysRevResearch.1.033012}, relying on a semiclassical approximation, in which the quantum oscillators are approximated by their classical counterparts in the weak quantum regime.

Nonetheless, measurement backaction itself alters dynamics of quantum trajectories, resulting in the emergence of a phase bias.
Hence, assessment of synchronization through a phase distribution hinges upon the choice of quadrature observables to be continuously measured.
In fact, we reported that continuous homodyne measurement of quadratures induces phase biases, even in the absence of synchronization in Ref.~\cite{PhysRevLett.132.093602}.
It is unclear which quadratures should be chosen and measured to representatively evaluate synchronization.
Furthermore, the quantum phase reduction approach defined the phase and limit-cycle solution according to a stochastic Schr\"{o}dinger equation associated with fixed quadratures, leading to incompatibility between cases where different quadratures are continuously monitored.

In this study, we extend the quantum phase reduction approach to heterodyne detection, which involves continuous measurement uniform over all possible quadratures.
Averaging out the measurement backaction for all quadrature angles, an unbiased quantum phase equation is derived from the stochastic Schr\"{o}dinger equation based on heterodyne detection.
Thus, the proposed method evaluates synchronization avoiding the measurement-induced phase bias.
For quantum van der Pol oscillators~\cite{PhysRevLett.111.234101}, we establish that the limit-cycle solution exhibits harmonic oscillation and that the measurement backaction induces no phase bias in heterodyne detection.
Moreover, our proposed method defines the limit-cycle solution, phase, and phase response curve independently of any specific quadratures due to the uniform continuous measurement.
By introducing these definitions from heterodyne to homodyne detection, the phase equation equally evaluates the measurement backaction for arbitrary quadratures.
Therefore, the proposed approach handles scenarios where the measured quadratures vary during the time evolution and where quantum oscillators are continuously monitored by different quadratures. 
Additionally, we demonstrate the efficacy of our method through numerical simulations, where uncoupled quantum oscillators subjected to common noise exhibit synchronization or yield phase clusters.
As a phenomenon not observed in classical synchronization, we reveal that the bosonic levels of these oscillators bound the possible number of the phase clusters between them; this limit is computed using the phase response curves.

\begin{figure}
    \centering
    \includegraphics[width=8.5cm]{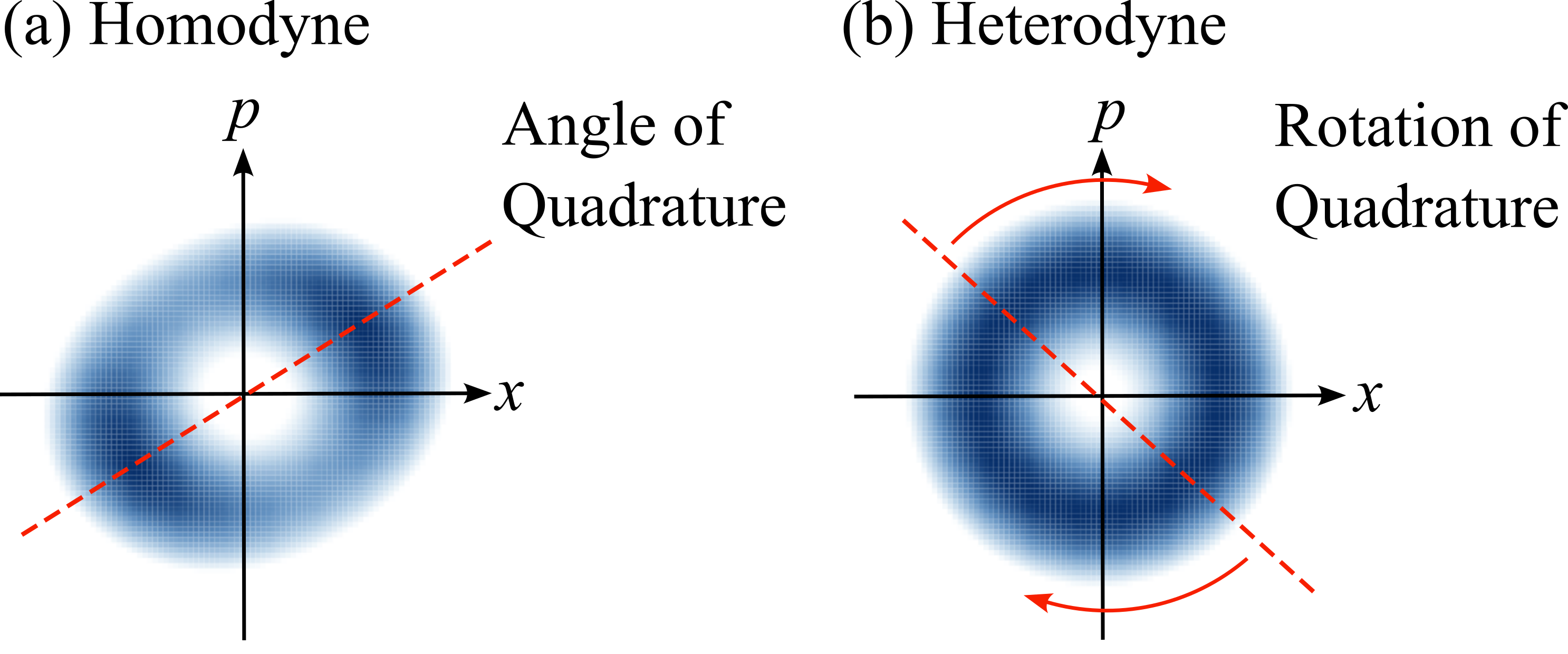}
    \caption{Distributions of quantum states in homodyne and heterodyne detection. (a) Heterodyne detection, where quadratures to be continuously monitored are fixed at certain angles, resulting in the biased distribution of quantum states due to their measurement backaction. (b) Heterodyne detection with quadratures rotating at sufficiently high frequency, allowing the unbiased distribution by averaging out measurement backaction across all angles of quadratures.}
    \label{fig:homovshetero}
\end{figure}
\section{METHODS}\label{sec:methods}
\subsection{Stochastic Schr\"{o}dinger equation for heterodyne detection}
\label{subsec:heteroSSE}
In open quantum systems, quantum limit-cycle oscillators are generally described by the Lindblad equation as follows~\cite{lindblad1976generators, breuer2002theory}:
\begin{align}
    \frac{d\rho}{dt}=-i[H,\rho]+\sum_{k=1}^K\mathcal{D}[L_k]\rho,
\label{eq:Lind}
\end{align}
where $H$ is Hamiltonian operator, $L_k$ are the jump operators and $\mathcal{D}[L]\rho$ is dissipator defined by $\mathcal{D}[L]\rho\equiv L\rho L^{\d}-(1/2)(L^{\d}L\rho+\rho L^{\d}L)$. 
However, the Lindblad equation describes the evolution of density operators of quantum systems, not that of the physically observable quantum states, known as quantum trajectory~\cite{carmichael2009open}. 
Quantum trajectories are the pure-state trajectories described by a stochastic Schr\"{o}dinger equation with the associated currents in the continuous measurement of the environment to which the quantum systems are coupled.
For continuous measurement, we use heterodyne detection for the quantum phase reduction in this paper, unlike our prior study based on homodyne detection.
Figure~\ref{fig:homovshetero} illustrates the difference between homodyne and heterodyne detection.
In homodyne detection, we select the angles for continuously measuring the quadratures corresponding to jump operators $L_k$, resulting in the measurement backaction biased according to these angles, as depicted in Fig.~\ref{fig:homovshetero}(a).
Conversely, for heterodyne detection, the measured quadratures are rotated at constant frequencies.
As the rotation reaches a sufficient frequency, the measurement backaction is uniformly distributed over all quadrature angles, as shown in Fig.~\ref{fig:homovshetero}(b).
\begin{table*}[ht]
\centering
\caption{Correspondence between classical and Lie-algebraic quantum phase reduction methods.}
\label{tab:phasereduction}
\begin{tabularx}{150mm}{LLLL}
    \quad & Classical & Quantum \\
    \hline \hline
    Space & Euclidean space & Hilbert space \\
    Differential equation & (Stochastic) differential equation & Stochastic Schr\"{o}dinger equation \\
    Phase response curve & Gradient & Lie algebra \\
    Perturbation & Vector & Unitary transformation \\
    \hline
\end{tabularx}
\end{table*}
Before introducing the quantum trajectory theory for heterodyne detection, we revisit that of homodyne detection, as the heterodyne stochastic Schr\"{o}dinger equation is generally derived from the homodyne counterpart~\cite{wiseman2009quantum}.
Two main approaches to implementation of homodyne detection exist: detection of stochastic homodyne currents by physical detectors~\cite{gardiner2004quantum} and continuous application of weak Gaussian measurements~\cite{jacobs2006straightforward}.
Quantum trajectories have been experimentally observed in various devices, such as cavities~\cite{guerlin2007progressive}, superconducting circuits~\cite{murch2013observing}, and mechanical resonators~\cite{PhysRevLett.123.163601}.
In homodyne detection, the quantum trajectories obey the following stochastic Schr\"{o}dinger equation in the Stratonovich form~\cite{Gisin_1992, gardiner2004quantum, barchielli2009quantum}:
\begin{align}
    d\ket{\psi}=&\Big[-iH+\sum_{k=1}^K-\frac{1}{2}(X_kL_k-\braket{X_kL_k})\nonumber\\
    &+\braket{X_k}(L_k-\braket{L_k})\Big]\ket{\psi}dt\nonumber\\
    &+\sum_{k=1}^K\left(L_k-\braket{L_k}\right)\ket{\psi}\circ dW_k,
    \label{eq:SSE}
\end{align}
where $\circ$ represents the Stratonovich calculus and $X_k\equiv L_k+L_k^{\d}$ is the quadrature corresponding to the jump operator $L_k$.
Furthermore, $\braket{O}$ represents the expectation value of $O$ in the state $\ket{\psi}$, that is, $\braket{O}\equiv \bra{\psi}O\ket{\psi}$. 
The random variables $dW_k$ are Wiener increments that satisfy $\mathbb{E}[dW_k]=0$ and $\mathbb{E}[dW_mdW_n]=\delta_{mn}dt$, where $\mathbb{E}[\cdot]$ represents the average over all possible quantum trajectories. 
The homodyne current $J_{\mathrm{hom},k}$ associated with the quadrature $X_k$ is defined as $J_{\mathrm{hom},k}\equiv \braket{X_k}+\xi_k(t)$, where $\xi_k(t)\equiv dW_k/dt$.
The average of all possible quantum trajectories described by Eq.~\eqref{eq:SSE}, that is, $\rho=\mathbb{E}[\ket{\psi}\bra{\psi}]$, satisfies the Lindblad equation [Eq.~\eqref{eq:Lind}].
It is noteworthy that a stochastic Schr\"{o}dinger equation satisfying the Lindblad equation is not unique, as the quadratures $X_k$ can be measured at any angles.
The rotation of the selected quadratures equals $U(1)$ transformation: $L_k\to L_k\exp(i\lambda_k)$, under which Eq.~\eqref{eq:Lind} remains invariant, and all forms of the stochastic Schr\"{o}dinger equations also satisfy Eq.~\eqref{eq:Lind} as well, where $\lambda_k$ is an arbitrary real parameter.
This transformation is physically implemented by rotation of the local oscillator signal~\cite{gardiner2004quantum}.

To average out the measurement backaction, we consider heterodyne detection in the fast rotation limit.
By rotating the local oscillators with constant frequencies $\Omega_k$, the jump operators $L_k$ also oscillate with such frequencies as: $L_k\to \mathrm{exp}(i\Omega_k t)L_k$.
As the frequency $\Omega_k$ increases to infinity, the backaction is averaged over the $U(1)$ transformation of $L_k$.
In the fast rotation limit, the stochastic Schr\"{o}dinger equation in heterodyne detection is derived from Eq.~\eqref{eq:SSE} as follows~\cite{wiseman2009quantum}:
\begin{align}
    d&\ket{\psi}\nonumber\\
    =&\left(-iH_{\mathrm{eff}}+\sum_{k=1}^K\frac{1}{2}\braket{L_k^{\d}L_k}+\braket{L_k^{\d}}(L_k-\braket{L_k})\right)\ket{\psi}dt \nonumber\\
    &+\sum_k(L_k-\braket{L_k})\ket{\psi}\circ d\tilde{W}_k^*,
    \label{eq:hetero}
\end{align}
where $H_{\mathrm{eff}}\equiv H-(i/2)\sum_kL_k^{\d}L_k$ is a non-Hermitian operator, called the effective Hamiltonian, and $d\tilde{W}_k\equiv(1/\sqrt{2})(dW_{k1}+idW_{k2})$ is the complex Wiener increment satisfying $\mathrm{E}[d\tilde{W}_k]=0$, $\mathrm{E}[d\tilde{W}_md\tilde{W}_n]=0$, and $\mathrm{E}[d\tilde{W}_m^*d\tilde{W}_n]=\delta_{m,n}dt$.
The measured heterodyne current is $J_{\mathrm{het},k}\equiv\braket{L_k}+\tilde{\xi}_k(t)$, where $\tilde{\xi}_k(t)\equiv d\tilde{W}_k/dt$.
It is easily shown that the heterodyne stochastic Schr\"{o}dinger equation [Eq.~\eqref{eq:hetero}] is independent of any specific quadratures, from the fact that Eq.~\eqref{eq:hetero} is invariant under arbitrary $U(1)$ transformation.

\subsection{Phase reduction based on heterodyne detection}
\label{subsec:PR_hetero}
Prior to discussing the specifics of its extension to heterodyne detection, let us briefly revisit the concept of Lie algebraic quantum phase reduction~\cite{PhysRevLett.132.093602}.
This method is a general framework for quantum synchronization, reducing multi-dimensional dynamics of quantum trajectories of the quantum limit-cycle oscillators, which fluctuate in the vicinity of a deterministic limit-cycle solution in the Hilbert space, into one-dimensional phase dynamics.
We separate the limit-cycle solution from the stochastic Schr\"{o}dinger equation and define the phase both on and outside this limit-cycle solution.
Additionally, the phase response curves, which represent the variation of the phase in response to sufficiently weak unitary transformations, are computed using Lie algebra framework.
Based on these definitions, the one-dimensional phase equation composed of the natural frequency of the limit-cycle solution and the phase response curves is derived from the stochastic Schr\"{o}dinger equation.
Table~\ref{tab:phasereduction} shows the correspondence between the conventional and quantum phase reduction methods.
Despite the difference between homodyne and heterodyne detection, the derivation process of the phase equation in this study follows that of the prior study~\cite{PhysRevLett.132.093602}.

To begin with, we define the limit-cycle solution for the stochastic Schr\"{o}dinger equation [Eq.~\eqref{eq:hetero}].
As an analog of classical cases~\cite{PhysRevLett.102.194102,PhysRevLett.98.184101,PhysRevLett.101.154101,GOLDOBIN2005126}, we propose to remove noise terms from a stochastic Schr\"{o}dinger equation in the Stratonovich form.
The resulting equation is defined as deterministic dynamics, which converges to a limit-cycle solution:
\begin{align}
    &d\ket{\psi}\nonumber\\
    &=\left(-iH_{\mathrm{eff}}+\sum_{k=1}^K\frac{1}{2}\braket{L_k^{\d}L_k}+\braket{L_k^{\d}}(L_k-\braket{L_k})\right)\ket{\psi}dt.
    \label{eq:limcyc}
\end{align}
Although numerous stochastic calculi exist, we define the limit-cycle solution solely based on the Stratonovich calculus, as it uniquely satisfies the norm preservation of the limit-cycle solution and the chain rule of differentiation~\cite{PhysRevLett.132.093602}.
To be defined as a physical trajectory like quantum trajectories, the limit-cycle solution should be a pure-state trajectory or satisfy norm preservation.
Equation~\eqref{eq:limcyc}, derived based on the Stratonovich calculus, satisfies $d\|\ket{\psi}\|=0$, where $\|\ket{\psi}\|\equiv\sqrt{\braket{\psi|\psi}}$, while the deterministic term of the stochastic Schr\"{o}dinger equation specified by any other stochastic calculus does not (see Appendix~\ref{appendix:norm preservation} for details). 
Additionally, the phase reduction approach generally requires a coordinate transformation between a state vector (corresponding to the quantum state $\ket{\psi}$ in this study) and a phase coordinate~\cite{doi:10.1080/00107514.2015.1094987}.
This transformation follows the chain rule of differentiation, which is uniquely satisfied in the Stratonovich calculus.
The state $\ket{\psi}$ converges to the limit-cycle solution $\ket{\psi_0}$ when Eq.~\eqref{eq:limcyc} satisfies the following condition,
\begin{align}
    \lim_{t\to\infty}|\braket{\psi(t)|\psi(t+T)}|=1,
    \label{eq:condition}
\end{align}
where $T$ denotes the period of the limit-cycle solution. 
Note that the state $\ket{\psi}$ is physically equivalent under the $U(1)$ transformation: $\ket{\psi} \to \exp{(i\lambda)}\ket{\psi}$.

Next, we define the phase for the quantum states in the Hilbert space using the limit-cycle solution $\ket{\psi_0}$. 
On the limit-cycle solution, we define the phase $\theta$ to change at constant frequency $\omega = 2\pi/T$ under Eq.~\eqref{eq:limcyc}. Furthermore, by employing the isochrone, the phase $\theta$ can be defined outside of the limit-cycle solution as follows:
\begin{align}
    \Theta[\ket{\psi(t)}]=\lim_{n\to \infty}\Theta[\ket{\psi(t+nT)}],
    \label{eq:isochrone}
\end{align}
where $\Theta[\ket{\psi}]$ represents the phase of the state $\ket{\psi}$ and the state $\ket{\psi}$ evolves according to Eq.~\eqref{eq:limcyc}.
Because the limit-cycle dynamics [Eq.~\eqref{eq:limcyc}] is invariant under the $U(1)$ transformation: $L_k \to \exp(i\lambda_k)L_k$, we can define the limit-cycle solution and phase regardless of any specific quadratures, unlike in our prior study based on homodyne detection. 

It should be noted that the phase response curves for quantum limit-cycle oscillators are calculated with respect to Lie algebra generators, rather than the basis vectors in the Euclidean space as in the conventional phase reduction approach~\cite{Kuramoto1984}. 
Arbitrary infinitesimal change of a pure state can be described by infinitesimal unitary transformation $U$, which can be decomposed into Lie algebra generators by the Taylor expansion as follows:
\begin{align}
    U&=\exp\left(\sum_{l=1}^{N^2-1}-ig_lE_l-ig_0I\right)\nonumber\\
    &\simeq I-\sum_{l=1}^{N^2-1}ig_lE_l-ig_0I,
    \label{eq:taylor}
\end{align}
where $I$ is an identity matrix, $E_l$ are Lie algebra generators of $SU(N)$, and real coefficients $g_l$ satisfies $|g_l|\ll 1$. 
The generators $E_l$ of $SU(N)$ are bases of traceless $N\times N$ Hermitian operators, for example, corresponding to Pauli matrices when $N=2$ and to Gell-Mann matrices when $N=3$ (see Appendix~\ref{appendix:GSN} for details).
Therefore, the phase response curve with respect to the generator $E_l$ is defined as follows:
\begin{align}
    Z_l(\theta)\equiv\lim_{g_l\to 0}\frac{\Theta[\exp(-ig_lE_l)\ket{\psi_0(\theta)}]-\Theta[\ket{\psi_0(\theta)}]}{g_l},
    \label{eq:PRC}
\end{align}
where $\ket{\psi_0(\theta)}$ represents the state $\ket{\psi}$ on the limit-cycle solution $\ket{\psi_0}$ with the phase $\theta$. 
Equation~\eqref{eq:PRC} represents a partial differentiation of the phase $\theta$ with respect to a unitary transformation by the generator $E_l$.
In the following, we assume that the perturbation is sufficiently weak such that the state $\ket{\psi}$ remains near the limit-cycle solution $\ket{\psi_0}$ and can be approximated by $\ket{\psi_0(\Theta[\ket{\psi}])}$.

Although the heterodyne stochastic Schr\"{o}dinger equation [Eq.~\eqref{eq:hetero}] is not explicitly presented as a linear equation of Hermitian operators, this equation can be decomposed into Lie algebra generators due to its nonlinearity, as we show below. 
Generally, a pure-state dynamics is described by a linear operator as follows:
\begin{align}
    d\ket{\psi}=-iB\ket{\psi}dt,
    \label{eq:-iA}
\end{align}
where $B$ is an arbitrary linear operator. 
Under Eq.~\eqref{eq:-iA}, derivative of the norm $\braket{\psi|\psi}$ is represented as follows:
\begin{align}
    d(\braket{\psi|\psi})&=(\bra{\psi}+d\bra{\psi})(\ket{\psi}+d\ket{\psi})-\braket{\psi|\psi}\nonumber\\
    &=d(\bra{\psi})\ket{\psi}+\bra{\psi}d(\ket{\psi})+d(\bra{\psi})d(\ket{\psi})\nonumber\\
    &=-i\braket{(B-B^{\d})}dt,
    \label{eq:-iA2}
\end{align}
where terms less than $O(dt)$ are neglected.
When $B$ is independent of the state $\ket{\psi}$, the norm preservation for arbitrary state $\ket{\psi}$ is satisfied only if $B$ is an Hermitian operator as $B=B^{\d}$.
On the other hand, when $B$ is dependent on the state $\ket{\psi}$, that is, when Eq.~\eqref{eq:-iA} is nonlinear, the norm preservation for arbitrary state $\ket{\psi}$ can be satisfied even if $B$ is not an Hermitian operator.
Indeed, the norm-preserved dynamics described by non-Hermitian operators, such as Eqs.~\eqref{eq:hetero} and \eqref{eq:limcyc}, can also be represented by an Hermitian operator as follows:
\begin{align}
\label{eq:A-braketA}
    -i\left[B(\ket{\psi})-\braket{B(\ket{\psi})}\right]\ket{\psi}dt&=-iH_B(\ket{\psi})\ket{\psi}dt,
\end{align}
where $H_B$ is the Hermitian operator
\begin{align}
    H_B(\ket{\psi})&\equiv \left[B(\ket{\psi})-\braket{B(\ket{\psi})}\right]\ket{\psi}\bra{\psi}+\mathrm{H.c.}
    \label{eq:H_B}
\end{align}
This statement holds true for the Stratonovich calculus.
It should be noted that the term $-i\braket{B}\ket{\psi}dt$, the difference between Eqs.~\eqref{eq:-iA} and \eqref{eq:A-braketA}, can be neglected since it corresponds to $U(1)$ transformation, which has no physical effect on the state $\ket{\psi}$.
Therefore, the heterodyne stochastic Schr\"{o}dinger equation [Eq.~\eqref{eq:hetero}] and Eq.~\eqref{eq:limcyc} can also be expressed as unitary transformations by the Hermitian operators $H_B(\ket{\psi})$.

To derive the phase equation from the heterodyne stochastic Schr\"{o}dinger equation [Eq.~\eqref{eq:hetero}], we focus on the stochastic terms of Eq.~\eqref{eq:hetero}, since the deterministic term of Eq.~\eqref{eq:hetero} corresponds merely to the natural frequency $\omega$ in the phase dynamics.
By following the procedure outlined in Eqs.~\eqref{eq:-iA}-\eqref{eq:H_B}, the stochastic terms of Eq.~\eqref{eq:hetero} can be represented by Hermitian operators as follows:
\begin{align}
    &(L_k-\braket{L_k})\ket{\psi}\circ d\tilde{W}_k^*\nonumber\\
    &=-iH_{k1}(\ket{\psi})\ket{\psi}\circ dW_{k1}-iH_{k2}(\ket{\psi})\ket{\psi}\circ dW_{k2},
    \label{eq:stochastic}
\end{align}
where $H_{k1}(\ket{\psi})$ and $H_{k2}(\ket{\psi})$ are the traceless Hermitian operators
\begin{align}
    \label{eq:Hk1}
    H_{k1}(\ket{\psi}) &\equiv \frac{i}{\sqrt{2}}(L_k-\braket{L_k})\ket{\psi}\bra{\psi}+\mathrm{H.c.},\\
    H_{k2}(\ket{\psi}) &\equiv \frac{1}{\sqrt{2}}(L_k-\braket{L_k})\ket{\psi}\bra{\psi}+\mathrm{H.c.}
    \label{eq:Hk2}
\end{align}
The stochastic terms of Eq.~\eqref{eq:hetero} are converted into the two orthogonal terms corresponding to the measurement backaction.
The traceless Hermitian operator $H_{km}$ can be decomposed into Lie algebra generators as $H_{km}=\sum_{l=1}^{N^2-1}g_{km,l}E_l$, where the coefficients $g_{km,l}$ are defined as $g_{km,l}\equiv\mathrm{Tr}[H_{km}E_l]$ due to trace orthogonality of Lie algebra generators.
Then, the following phase equation is derived according to the chain rule of differentiation.
\begin{align}
    \frac{d\theta}{dt}= \omega +\sum_{k=1}^K\sum_{m=1}^2\sum_{l=1}^{N^2-1}Z_l(\theta)g_{km,l}(\theta)\circ \xi_{km}(t),
    \label{eq:PE_str}
\end{align}
where $g_{km,l}(\theta)$ is evaluated at $\ket{\psi}=\ket{\psi_0(\theta)}$ on the limit-cycle solution.
The phase equation~\eqref{eq:PE_str} in the Stratonovich form can be converted into an equivalent equation in the Ito form (see Appendix~\ref{appendix:stochastic calculus} for details).
\begin{align}
    \frac{d\theta}{dt}&=\omega+\frac{1}{2}\sum_{k=1}^K\sum_{m=1}^2\frac{dY_{km}(\theta)}{d\theta}Y_{km}(\theta)\nonumber\\
    &+\sum_{k=1}^K\sum_{m=1}^2 Y_{km}(\theta)\xi_{km}(t),
    \label{eq:PE_ito}
\end{align}
where $Y_{km}(\theta)\equiv\sum_{l=1}^{N^2-1}Z_l(\theta)g_{km,l}(\theta)$.

\begin{figure}
\includegraphics[width=8.5cm]{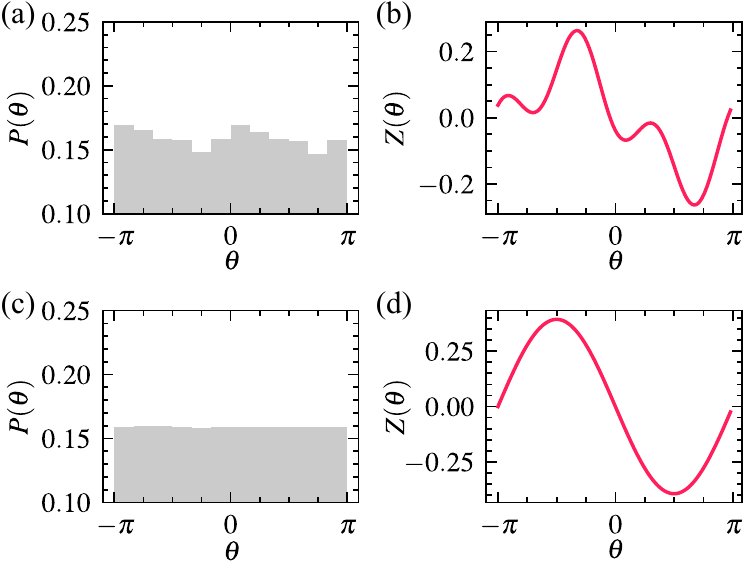}
\caption{
Phase distribution $P(\theta)$ in steady state and phase response curve $Z(\theta)$ of quantum van der Pol oscillators in homodyne and heterodyne detection. (a) and (c) Phase distribution in (a) heterodyne detection and (c) homodyne detection. (b) and (d) phase response curve with respect to a harmonic drive in (b) heterodyne detection and (d) homodyne detection. The gray histograms are computed from simulations of the stochastic Schr\"{o}dinger equation. The parameters are $(\Delta$, $\gamma_{g}$, $\gamma_{d})=$$(1.0$, $0.2$, $1.0)$. The phase response curve with respect to the harmonic drive is calculated for an Hermitian operator $i(a-a^{\d})$.
}\label{fig:hetero}      
\end{figure}
To demonstrate that the phase equation [Eq.~\eqref{eq:PE_str}] is unbiased and independent of specific quadratures, we consider the case of a quantum van der Pol oscillator in heterodyne detection as an example.
The Lindblad equation of the quantum van der Pol oscillator is represented as follows~\cite{PhysRevLett.111.234101}:
\begin{align}
    \frac{d\rho}{dt} = -i[\Delta a^{\d}a,\rho]+\gamma_{g}\mathcal{D}[a^{\d}]\rho+\gamma_{d}\mathcal{D}[a^2]\rho,
    \label{eq:qvdP}
\end{align}
where $\Delta$ is the detuning between the oscillator and the frequency of a rotating frame, $a$ and $a^{\d}$ are annihilation and creation operators, respectively, and $\gamma_{g}$ and $\gamma_{d}$ are the jump rates corresponding to one-particle gain and two-particle loss, respectively. 
Owing to its $U(1)$ symmetry, we establish that, if it exists, the limit-cycle solution of the quantum van der pol oscillators obtained according to Eq.~\eqref{eq:limcyc} is always harmonic oscillation; the limit-cycle solution is not distorted at any specific angles of the measured quadratures (see Appendix~\ref{appendix:hetero cyc} for details).
Based on this limit-cycle solution, we calculate the phase response curves for each jump operator, by following Eqs.~\eqref{eq:stochastic}-\eqref{eq:Hk2}; for both jump operators, they correspond to first- and second-harmonic orthogonal trigonometric functions, respectively.
According to Eq.~\eqref{eq:PE_str}, we derive the phase equation corresponding to the quantum van der Pol model as follows:
\begin{align}
    \frac{d\theta}{dt}&=\omega+v_{1}\sin{\theta}\circ \xi_{11}+v_{1}\cos{\theta}\circ \xi_{12}\nonumber\\
    &+v_{2}\sin{2\theta}\circ \xi_{21}+v_{2}\cos{2\theta}\circ \xi_{22}\nonumber\\
    &=\omega + v \xi,
    \label{eq:PE_qvdP}
\end{align}
where $v_1$ and $v_2$ are the backaction strengths corresponding to one-particle gain and two-particle loss, respectively, and $v\equiv\sqrt{{v_1}^2+{v_2}^2}$. 
The transition from the first to the second line utilizes the relation $\sum_{k=1}^K v_k dW_k=\sqrt{\sum_{k=1}^K v_k^2}dW$.
Consequently, in the phase equation, the averaged backaction in heterodyne detection is simple white noise and serves as constant phase diffusion, while it induces clustering in homodyne detection (see Appendix~\ref{appendix:hetero PRC} for details).
Figure~\ref{fig:hetero} illustrates the phase distribution in the steady state and the phase response curve with respect to the harmonic drive both in homodyne and heterodyne detection. 
In homodyne detection, clustering occurs in the steady state even without external drive in Fig.~\ref{fig:hetero}(a); however, no clustering is observed in heterodyne detection, as depicted in Fig.~\ref{fig:hetero}(c).
Furthermore, while the previous method based on homodyne detection demonstrates a distorted phase response curve in Fig.~\ref{fig:hetero}(b), the proposed method for heterodyne detection shows an undistorted sinusoidal wave in Fig.~\ref{fig:hetero}(d).
This can be attributed to the fact that, in the proposed method, the phase is independent of the quadratures and symmetric against rotation.

\subsection{Application of phase reduction based on heterodyne detection to homodyne detection}
\label{subsec:PR_homo}
As the definition of the phase depends on the fixed quadratures, Lie algebraic phase reduction based on homodyne detection~\cite{PhysRevLett.132.093602} does not apply to the scenario where the measured quadratures are modulated.
To address this issue, we apply the definition of the phase in the proposed method for heterodyne detection, which is independent of any quadratures, to homodyne detection.
This application does not imply the introduction of an additional phase reduction approach specifically for homodyne detection, distinct from the one proposed in Subsec.~\ref{subsec:PR_hetero} for heterodyne detection.
\begin{figure}
    \centering
    \includegraphics[width=8.5cm]{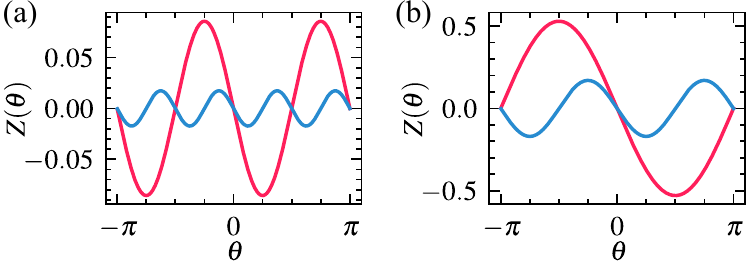}
    \caption{Phase response curves of quantum van der Pol oscillator in homodyne detection. Phase response curves with respect to the measurement backaction corresponding to $L_1=\sqrt{\gamma_1}a^{\dagger}$ (red line) and $L_2 = \sqrt{\gamma_2}a^2$ (blue line) for (a) deterministic and (b) stochastic terms. The parameters are $(\Delta$, $\gamma_{g}$, $\gamma_{d})=$$(1.0$, $0.2$, $1.0)$.}
    \label{fig:diff_homo_hetero}
\end{figure}

To derive the phase equation from the homodyne stochastic Schr\"{o}dinger equation [Eq.~\eqref{eq:SSE}] based on the proposed method, we focus on the difference between Eq.~\eqref{eq:SSE} and the limit-cycle dynamics [Eq.~\eqref{eq:limcyc}] as follows:
\begin{align}
    \label{eq:diff}
    d\ket{\psi}=& \Big[\sum_{k=1}^{K}-\frac{1}{2}(L_k^2-\braket{L_k^2})+\braket{L_k}(L_k-\braket{L_k})\Big]\ket{\psi}dt\nonumber\\
    &+(L_k-\braket{L_k})\ket{\psi}\circ dW_k
\end{align}
Because Eq~\eqref{eq:diff} also satisfies the norm preservation $d\|\ket{\psi}\|=0$, it can be converted into the phase equation by following the procedure of Eqs.~\eqref{eq:-iA}-\eqref{eq:H_B}, explained in Subsec.~\ref{subsec:PR_hetero}.
Figure~\ref{fig:diff_homo_hetero} illustrates the phase response curves corresponding to the deterministic and stochastic terms of Eq.~\eqref{eq:diff} for each jump operator.
It is evident from these phase response curves that the measurement backaction induces the phase bias in homodyne detection, consistent with the fact revealed in Ref.~\cite{PhysRevLett.132.093602}.
Nevertheless, since the definition of the phase does not rely on any quadratures, the phase response curves show waveform without distortion, even in homodyne detection, unlike that of the prior approach as shown in Fig.~\ref{fig:diff_homo_hetero}(b).
Hence, the proposed method can represent the homodyne measurement backaction and the phase bias induced by it, while the phase is defined regardless of this bias.
Applying the $U(1)$ transformation: $L_k \to \exp(i\lambda_k)L_k$ to Eq.~\eqref{eq:diff}, cases measuring any other quadratures can also be considered in the same way. 
Thus, by employing the definitions introduced in Subsec.~\ref{subsec:PR_hetero}, the proposed method applies to arbitrary cases of homodyne detection, including the cases where the angle, at which the quadratures are measured, is controlled via feedback, or where multiple quantum oscillators are continuously measured with different quadratures.

\section{EXAMPLES}\label{sec:examples}
Owing to the unbiased nature of the phase in our proposed method, phase synchronization dynamics between quantum oscillators can be analyzed without considering measurement-induced phase clusters.
To demonstrate the effectiveness of our proposed method, we conduct numerical simulations wherein synchronization occurs between two uncoupled quantum oscillators subjected to common Hamiltonian noise.
Noise-induced synchronization refers to the synchronization of quantum oscillators subjected to common noise, even without coupling between them.
This phenomenon has been observed for various physical systems, such as uncoupled neurons~\cite{PhysRevLett.88.138103} and single mode $\mathrm{CO_2}$ lasers~\cite{PhysRevE.67.066220}.
Theoretically, noise-induced synchronization has been analyzed in various nonlinear oscillator models by applying the phase reduction method~\cite{PhysRevLett.98.184101} and Lyapunov exponent~\cite{doi:10.1063/1.1493096}. 
Furthermore, under suitable conditions, perfect synchronization between the oscillators can also be caused by common noise~\cite{PhysRevLett.98.184101}.
Recently, Schmolke and Lutz reported that a spin chain subjected to Hermitian noise synchronize~\cite{PhysRevLett.129.250601}.
In this study, we numerically simulate the case of uncoupled quantum limit-cycle oscillators subjected to common noise synchronize under continuous measurement.

Let us consider the case in which two quantum van der Pol oscillators are subjected to a Hermitian noise $H_N$
\begin{align}
    \frac{d\rho_1}{dt} &= -i[\Delta a_1^{\d}a_1+H_N,\rho]+\gamma_{g}\mathcal{D}[a_1^{\dagger}]\rho+\gamma_{d}\mathcal{D}[a_1^2]\rho,\nonumber\\
    \frac{d\rho_2}{dt} &= -i[\Delta a_2^{\d}a_2+H_N,\rho]+\gamma_{g}\mathcal{D}[a_2^{\dagger}]\rho+\gamma_{d}\mathcal{D}[a_2^2]\rho,\nonumber\\
    H_N &\equiv S_N\circ \xi_N(t),
    \label{eq:Lind_common}
\end{align}
where $S_N$ is an Hermitian operator; in this simulation, we specify the cases $S_N=i(a-a^{\dagger})$ and $S_N=i(a^2-a^{\dagger2})$.
In physical context, the Hermitian noise is caused by an ac-Stark shift~\cite{PhysRevLett.122.050501} or a thermal bath~\cite{PhysRevB.96.115408}. 
To focus on the noise-induced synchronization without the phase bias caused by the measurement backaction, we use heterodyne detection for the quantum phase reduction.
The Hamiltonian noise $H_N$ can be decomposed into Lie algebra generators as $H_N=\sum_{l=1}^{N^2-1}g_{N,l}E_l$. 
According to Eq.~\eqref{eq:PE_qvdP}, the phase equations corresponding to Eq.~\eqref{eq:Lind_common} can be derived as follows:
\begin{align}
    \frac{d\theta_1}{dt}&=\omega + v\circ \xi_1 + Z_{H_N}(\theta_1)\circ \xi_N,\nonumber\\
    \frac{d\theta_2}{dt}&=\omega + v\circ \xi_2 + Z_{H_N}(\theta_2)\circ \xi_N,
    \label{eq:PE_common}
\end{align}
where $Z_{H_N}(\theta)\equiv\sum_{l=1}^{N^2-1}g_lZ_l(\theta)$ is the phase response curve with respect to the Hamiltonian noise $H_N$.

To theoretically analyze synchronization dynamics between the two oscillators, we consider the slow limit, where the phase difference between them varies much more slowly than their frequency $\omega$. 
Then, we highlight the relative phase $\phi_i\equiv \theta_i-\omega t$, which represents a disturbance from its natural frequency $\omega$. 
Equation~\eqref{eq:PE_common} is described with respect to the relative phase $\phi$ as follows:
\begin{align}
    \frac{d\phi_1}{dt}&=v\circ \xi_1 + Z_{H_N}(\phi_1 + \omega t)\circ \xi_N,\nonumber\\
    \frac{d\phi_2}{dt}&=v\circ \xi_2 + Z_{H_N}(\phi_2 + \omega t)\circ \xi_N,
    \label{eq:PE_common2}
\end{align}
The joint probability distribution $P(\phi_1,\phi_2,t)$ is governed by the Fokker-Planck equation~\cite{risken1996fokker} as follows:
\begin{align}
\label{eq:common_FPE1}
    \frac{\partial P}{\partial t}&=\sum_{i=1}^2\frac{dZ_{H_N}(\phi_i+\omega t)}{d\phi_i}Z_{H_N}(\phi_i+\omega t)+\sum_{i,j}\frac{\partial^2}{\partial \phi_i \partial \phi_j}D_{ij},\\
    D_{ij}&\equiv \frac{1}{2}p^2\delta_{ij}+\frac{1}{2}Z_{H_N}(\phi_i+\omega t)Z_{H_N}(\phi_j +\omega t),
    \label{eq:common_FPE2}
\end{align}
In the slow limit, we can average the drift and diffusion coefficients in Eqs.~\eqref{eq:common_FPE1} and \eqref{eq:common_FPE2} over the period $T$ while keeping the phase difference $\phi_{-}\equiv\phi_1-\phi_2$ constant (in the following, we adopt the procedure in Ref.~\cite{Kuramoto1984,PhysRevLett.98.184101}).
\begin{align}
    \tilde{D}_{ij}(\phi_{i}-\phi_{j})&\equiv\frac{1}{T}\int_0^{t+T} D_{ij}dt,\nonumber\\
    &=h(\phi_{i}-\phi_{j})+v^2 \delta_{ij},
    \label{eq:averagedD}
\end{align}
where the correlation function $h(\phi_{-})$ is defined as follows:
\begin{align}
    h(\phi_{-})\equiv\frac{1}{2\pi}\int_0^{2\pi}Z_{H_N}(\phi_{-}+\phi')Z_{H_N}(\phi')d\phi'.
    \label{eq:correlation}
\end{align}
The correlation function $h(\phi_{-})$ has maximum value at $\theta=0$ as long as $Z_{H_N}(\theta)$ is smooth. 
By averaging the Fokker-Planck equation over the period $T$, the drift terms vanish. 
As shown in Eq.~\eqref{eq:averagedD}, the diffusion coefficient depends only on the phase difference $\phi_-$.
Then, substituting Eqs.~\eqref{eq:averagedD} and \eqref{eq:correlation} to Eq.\eqref{eq:common_FPE1}, we obtain the Fokker-Planck equation averaged over period $T$ as follows:
\begin{align}
    \frac{\partial}{\partial t}P(\phi_1,\phi_2,t)=&\frac{1}{2}[h(0)+v^2]\left\{\left(\frac{\partial}{\partial\phi_1}\right)^2+\left(\frac{\partial}{\partial\phi_2}\right)^2\right\}P\nonumber\\
    &+\frac{\partial^2}{\partial \phi_1 \partial \phi_2}[h(\phi_{-})P].
    \label{eq:commonFPE2}
\end{align}
Moreover, to see the phase dynamics in detail, we transform the phase variables $\phi_1$ and $\phi_2$ to $\phi_{+}\equiv\phi_1+\phi_2$ and $\phi_{-}$. The probability distribution $Q(\phi_-)$ is defined as $Q(\phi_{-},t)\equiv\int_0^{2\pi}P(\phi_1,\phi_2,t)d\phi_{+}$ and the corresponding Fokker-Planck equations are represented as follows:
\begin{align}
    \frac{\partial Q(\phi_-)}{\partial t}&=\frac{\partial^2}{\partial\phi_{-}^2}\{[h(0)-h(\phi_{-})]+v^2\}Q(\phi_{-},t).
    \label{eq:commonFPE3}
\end{align}
Clearly, Eq.~\eqref{eq:commonFPE3} has a steady state solution when $Q(\phi_{-})$ is represented as
\begin{align}
    Q(\phi_{-})=\frac{q_0}{[h(0)-h(\phi_{-})]+v^2},
    \label{eq:steady}
\end{align}
where $q_0$ is a normalization constant. 
\begin{figure}
\includegraphics[width=8.5cm]{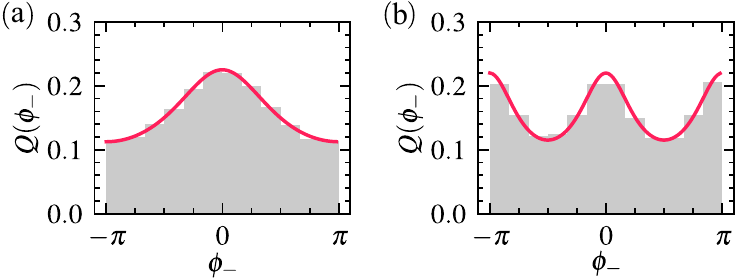}
\caption{Phase-difference distribution of quantum van der Pol oscillators subjected to common noise in steady state.
(a) and (b) phase-difference distribution $Q(\phi_{-})$ in the steady state in heterodyne detection subjected to the common noise. The gray histograms are computed from simulations of the stochastic Schr\"{o}dinger equation and the red lines are computed from simulations of the phase equation. The parameters are $(\Delta$, $\gamma_{g}$, $\gamma_{d})=$$(1.0$, $0.2$, $1.0)$. The common Hermitian noise $S_N$ is $S_N=i(a-a^{\d})$ for (a), $S_N=i(a^2-a^{\dagger 2})$ for (b).}
\label{fig:common}      
\end{figure}

Although expression of the phase-difference distribution $Q(\phi_{-})$ in steady state [Eq.~\eqref{eq:steady}] itself is the same as that of the classical noise-induced synchronization~\cite{PhysRevLett.98.184101}, we can identify a distinction between the classical and quantum cases: the number of possible phase clusters between the two quantum oscillators is limited by their bosonic levels.
In classical cases, considering common noise with higher-order terms allowed for an arbitrarily large number of phase clusters between two oscillators.
As shown in Eq.~\eqref{eq:steady}, the phase distribution in the steady state depends on the correlation function $h(\phi_{-})$.
For the quantum van der Pol oscillators in heterodyne detection, the phase response curves are given as trigonometric functions with the modes of $N-1$ or less. 
For instance, when $S_N=i(a-a^{\dagger})$ or $i(a^2-a^{\dagger 2})$, the corresponding phase response curve is a first or second harmonic, resulting in a synchronized state or cluster state in steady state as shown in Figs.~\ref{fig:common}(a) and (b).
The correlation function $h(\phi_{-})$ can be represented by a linear combination of the trigonometric functions with modes of $N-1$ or less.
Hence the number of possible phase clusters in response to an external perturbation is restricted by their bosonic levels, unlike the classical case.

\section{CONCLUSION}
In this study, we extend Lie algebraic quantum phase reduction~\cite{PhysRevLett.132.093602} to heterodyne detection, where all quadrature observables are uniformly measured.
Averaging out the measurement backaction, the proposed method derives an unbiased phase equation in the absence of measurement-induced phase clusters.
Moreover, our proposed method also applies to arbitrary cases of homodyne detection, including the scenario where the quadratures are modified during the time evolution.
Using the proposed method, we conduct numerical simulations where uncoupled quantum van der Pol oscillators subjected to common noise exhibit synchronization.
Recently, various phenomena intrinsic to quantum oscillators have been reported in field of quantum synchronization, spurring the demand for their analysis~\cite{PhysRevLett.117.073601,PhysRevLett.118.243602,PhysRevA.108.022216}.
In studies of classical synchronization, the extensions of phase reduction theory have been developed to increase its effectiveness for complex synchronization dynamics~\cite{PhysRevLett.111.214101,PhysRevE.101.022220}.
Incorporating these techniques to Lie algebraic quantum phase reduction may enhance its adaptability to the synchronization phenomena specific to quantum oscillators.

\appendix
\section{Norm preservation of limit-cycle equation}\label{appendix:norm preservation}
To ensure the limit-cycle solution of the stochastic Schr\"{o}dinger equation to be a pure-state trajectory, its deterministic term must satisfy the norm preservation condition $\|\ket{\psi}\|=1$.
However, it should be noted that the deterministic terms of the stochastic differential equations varies with the choice of stochastic calculus for their evaluation.
We establish that the Stratonovich calculus uniquely satisfies the norm preservation of the limit-cycle dynamics.
Considering an arbitrary stochastic calculus, we evaluate the stochastic Schr\"{o}dinger equation at an arbitrary time $t+pdt$ during $[t,t+dt]$, where $0\leq p\leq 1$ ($p=0$ in the Ito calculus and $p=1/2$ in the Stratonovich calculus).
When the stochastic Schr\"{o}dinger equation is evaluated at time $t+pdt$, it is represented as follows:
\begin{align}
    &d\ket{\psi(t)}\nonumber\\
    &=\left[-iH_{\mathrm{eff}}+\sum_{k=1}^K \braket{L_k^{\d}}L_k - \frac{1}{2}\braket{L_k^{\d}}\braket{L_k}\right]\ket{\psi(t+pdt)}dt\nonumber\\
    &+p\left[\braket{L^{\d}_kL_k}-\braket{L_k^{\d}}\braket{L_k}\right]\ket{\psi(t+pdt)}dt\nonumber\\
    &+(L_k-\braket{L_k})\ket{\psi(t+pdt)}dZ^*(t),
    \label{eq:appendix:SSE_p}
\end{align}
where terms less than $O(dt)$ are neglected. The time derivative of the norm $\|\ket{\psi}\|$ under the deterministic terms of Eq.~\eqref{eq:appendix:SSE_p} is described as follows:
\begin{align}
    &d(\braket{\psi|\psi})\nonumber\\
    &=(\bra{\psi}+d\bra{\psi})(\ket{\psi}+d\ket{\psi})-\braket{\psi|\psi}\nonumber\\
    &=(d\bra{\psi})\ket{\psi}+\bra{\psi}(d\ket{\psi})+(d\bra{\psi})(d\ket{\psi})\nonumber\\
    &=(2p-1)\sum_{k=1}^K\left[\braket{L_k^{\d}L_k}-\braket{L_k^{\d}}\braket{L_k}\right]dt,
\end{align}
Clearly, the norm preservation $d\|\ket{\psi}\|=0$ for an arbitrary state $\ket{\psi}$ is satisfied only if $p=1/2$, the Stratonovich calculus.

\section{Generators of $SU(N)$}\label{appendix:GSN}
The generators of $SU(N)$ correspond to generalized Gell-Mann matrices.
The generalized Gell-Mann matrices are composed of off-diagonal matrices $f_{k,l}$ and diagonal matrices $h_{k,l}$.
\begin{align}
    \Lambda_{j,k} &\equiv
\begin{cases}
     \text{for $1\leq j<k\leq N$}\\
     O_{k,j}+O_{j,k},\\
    \text{for $1\leq k<j \leq N$}\\
    -i(O_{j,k}-O_{k,j}),\\
    \text{for $1\leq j=k \leq N-1$}\\
     \sqrt{\frac{2}{j(j+1)}}\left(\sum_{l=1}^j O_{l,l}-jO_{j+1,j+1}\right),\\
     \end{cases}
\end{align}
where $O_{j,k}$ is the matrix with $1$ in the $jk$-th entry and $0$ elsewhere. The generators $E_{j,k}$ are defined so that they are normalized with respect to the trace norm:
\begin{align}
    E_{j,k}&\equiv \frac{1}{\sqrt{\mathrm{Tr}[\Lambda_{j,k}^2]}}\Lambda_{j,k}\nonumber,\\
\end{align}

\section{Conversion between Ito and Stratonovich calculi}\label{appendix:stochastic calculus}
Typically, a stochastic differential equation is represented by two different forms, the Ito and Stratonovich forms, which are given as follows:
\begin{align}
    f(t)dW(t)&\equiv f(t)\left(W(t+dt)-W(t)\right),\\
     f(t)\circ dW(t)&\equiv f\left(t+\frac{dt}{2}\right)\left(W(t+dt)-W(t)\right),
\end{align}
where $W(t)$ is a Wiener process. In the Stratonovich interpretation, a calculus is performed at the midpoint of the interval $[t, t+dt]$. Moreover, the two forms can be converted to each other using the following transformation:
\begin{align}
    f(t)\circ dW(t)=f(t)dW(t)+\frac{1}{2}df(t)dW(t).
\label{eq:conversion}
\end{align}
The conversion of Eq.~\eqref{eq:conversion} is performed according to the Ito rule:
\begin{align}
    dW(t)^2&=dt,\\
    dW(t)dt&=0.
\end{align}

\section{Limit-cycle solution of quantum van der Pol oscillator in heterodyne detection}\label{appendix:hetero cyc}
To show that the limit-cycle solution of the quantum van der Pol oscillator is a harmonic oscillation in heterodyne detection, we provide a brief proof for it.
We define a unitary transformation $R(\lambda)$ corresponding to harmonic oscillation as $R(\lambda)\equiv\exp(-i a^{\d}a \lambda)$.
According to Eq.~\eqref{eq:limcyc}, dynamics of limit cycle for quantum van der Pol oscillators~[Eq.~\eqref{eq:qvdP}] is represented as follows:
\begin{align}
    d\ket{\psi}=&\big[-ia^{\d}a-\frac{1}{2}(\gamma_gaa^{\d}+\gamma_da^{\d2}a^2)+\gamma_g\braket{a}(a^{\d}-\braket{a^{\d}})\nonumber\\
    &+\gamma_d\braket{a^{\d2}}(a^2-\braket{a^2})\big]\ket{\psi}dt.
    \label{eq:appendix:qvdP_limcyc}
\end{align}
By performing this unitary transformation on the quantum state $\ket{\psi}$ in Eq.~\eqref{eq:appendix:qvdP_limcyc} of quantum van der Pol oscillators, we find $R(\ket{\psi}+d\ket{\psi})=R(\lambda)\ket{\psi}+d(R(\lambda)\ket{\psi})$, that is, time evolution in Eq.~\eqref{eq:appendix:qvdP_limcyc} and the unitary transformation $R(\lambda)$ commute.
By introducing the operator $C(\beta)$ as $C(\beta)\ket{\psi(t)}=\ket{\psi(t+\beta)}$ under Eq.~\eqref{eq:appendix:qvdP_limcyc}, it is clear $C(\beta)R(\lambda)\ket{\psi}=R(\lambda)C(\beta)\ket{\psi}$.
Using this relation in the limit-cycle solution $\ket{\psi_0(\theta)}$, we can derive
\begin{align}
    C(\beta)R(\lambda)\ket{\psi_0(\theta)}&=R(\lambda)C(\beta)\ket{\psi_0(\theta)}\nonumber\\
    &=R(\lambda)\ket{\psi_0(\theta+\omega\beta)}.
    \label{eq:appendix:hetero_cyc}
\end{align}
To take $\beta \to \infty$, the left-hand side of Eq.~\eqref{eq:appendix:hetero_cyc} converges on the limit-cycle solution $\ket{\psi_0}$. 
Thus, the limit-cycle solution $\ket{\psi_0}$ is a harmonic oscillation and is described as an circle on $xp$ plane.

\section{Phase response curve of quantum van der Pol oscillator in heterodyne detection}\label{appendix:hetero PRC}
We prove that in heterodyne detection, the phase response curves corresponding to jump operators in the quantum van der Pol oscillators are constant white noise.
Unitary transformation $R(\lambda)=\exp(i \lambda a^{\d}a)$ and annihilation and creation operators satisfy $R(\lambda)a=e^{i\lambda}aR(\lambda)$.
Using this fact, when we define the derivative of quantum states with respect to perturbations as $d\ket{\psi}=(a^{\d}-\braket{a^{^\d}})\ket{\psi}dt$, we can derive
\begin{align}
    d[R(\lambda)\ket{\psi}]&=e^{-i\lambda}d\ket{\psi}\nonumber\\
    &=(a^{\d}e^{-i\lambda}-\braket{a^{\d}e^{-i\lambda}})\ket{\psi}dt.
    \label{eq:appendix:rot_perturb}
\end{align}
According to Appendix~\ref{appendix:hetero cyc}, unitary transformation $R(\lambda)$ and dynamics of limit cycle defined by Eq.~\eqref{eq:limcyc} commute.
Employing this and Eq.~\eqref{eq:appendix:rot_perturb}, we derive $Z_r(\lambda)=\cos(\lambda)Z_r(0)-\sin(\lambda) Z_i(0)$ where $Z_r(\theta)$ and $Z_i(\theta)$ are phase response curves with respect to perturbations of $a^{\d}$ and $ia^{\d}$, respectively.
Then, we find that $Z_r(\theta)$ can be represented as $Z_r(\theta)=c\cos(\theta+\theta_0)$ where $\theta_0$ is constant.
Considering the case where $\lambda = \pi/2$ in Eq.~\eqref{eq:appendix:rot_perturb}, it can be easily deduced that $Z_r(\lambda)=-Z_i(\lambda-\pi/2)$.
Finally, we can obtain $\sqrt{[Z_r(\theta)dW_1]^2+[Z_i(\theta)dW_2]^2}=\sqrt{2}c dW$.
A similar discussion can be applied to a jump operator $a^2$.

\begin{acknowledgments}
This work was supported by JSPS KAKENHI Grant No. JP22H03659.
\end{acknowledgments}

\begin{thebibliography}{55}%
\makeatletter
\providecommand \@ifxundefined [1]{%
 \@ifx{#1\undefined}
}%
\providecommand \@ifnum [1]{%
 \ifnum #1\expandafter \@firstoftwo
 \else \expandafter \@secondoftwo
 \fi
}%
\providecommand \@ifx [1]{%
 \ifx #1\expandafter \@firstoftwo
 \else \expandafter \@secondoftwo
 \fi
}%
\providecommand \natexlab [1]{#1}%
\providecommand \enquote  [1]{``#1''}%
\providecommand \bibnamefont  [1]{#1}%
\providecommand \bibfnamefont [1]{#1}%
\providecommand \citenamefont [1]{#1}%
\providecommand \href@noop [0]{\@secondoftwo}%
\providecommand \href [0]{\begingroup \@sanitize@url \@href}%
\providecommand \@href[1]{\@@startlink{#1}\@@href}%
\providecommand \@@href[1]{\endgroup#1\@@endlink}%
\providecommand \@sanitize@url [0]{\catcode `\\12\catcode `\$12\catcode
  `\&12\catcode `\#12\catcode `\^12\catcode `\_12\catcode `\%12\relax}%
\providecommand \@@startlink[1]{}%
\providecommand \@@endlink[0]{}%
\providecommand \url  [0]{\begingroup\@sanitize@url \@url }%
\providecommand \@url [1]{\endgroup\@href {#1}{\urlprefix }}%
\providecommand \urlprefix  [0]{URL }%
\providecommand \Eprint [0]{\href }%
\providecommand \doibase [0]{https://doi.org/}%
\providecommand \selectlanguage [0]{\@gobble}%
\providecommand \bibinfo  [0]{\@secondoftwo}%
\providecommand \bibfield  [0]{\@secondoftwo}%
\providecommand \translation [1]{[#1]}%
\providecommand \BibitemOpen [0]{}%
\providecommand \bibitemStop [0]{}%
\providecommand \bibitemNoStop [0]{.\EOS\space}%
\providecommand \EOS [0]{\spacefactor3000\relax}%
\providecommand \BibitemShut  [1]{\csname bibitem#1\endcsname}%
\let\auto@bib@innerbib\@empty
\bibitem [{\citenamefont {Kuramoto}(1984)}]{Kuramoto1984}%
  \BibitemOpen
  \bibfield  {author} {\bibinfo {author} {\bibfnamefont {Y.}~\bibnamefont
  {Kuramoto}},\ }\href@noop {} {\emph {\bibinfo {title} {Chemical Oscillations,
  Waves, and Turbulence}}}\ (\bibinfo  {publisher} {Springer},\ \bibinfo
  {address} {New York},\ \bibinfo {year} {1984})\BibitemShut {NoStop}%
\bibitem [{\citenamefont {Winfree}(2001)}]{winfree2001geometry}%
  \BibitemOpen
  \bibfield  {author} {\bibinfo {author} {\bibfnamefont {A.}~\bibnamefont
  {Winfree}},\ }\href@noop {} {\emph {\bibinfo {title} {The Geometry of
  Biological Time}}},\ \bibinfo {edition} {2nd}\ ed.\ (\bibinfo  {publisher}
  {Springer, New York},\ \bibinfo {year} {2001})\BibitemShut {NoStop}%
\bibitem [{\citenamefont {Izhikevich}(2007)}]{izhikevich2007dynamical}%
  \BibitemOpen
  \bibfield  {author} {\bibinfo {author} {\bibfnamefont {E.}~\bibnamefont
  {Izhikevich}},\ }\href@noop {} {\emph {\bibinfo {title} {Dynamical Systems in
  Neuroscience}}}\ (\bibinfo  {publisher} {MIT Press},\ \bibinfo {address}
  {Cambridge, Massachusetts},\ \bibinfo {year} {2007})\BibitemShut {NoStop}%
\bibitem [{\citenamefont {Strogatz}(2018)}]{strogatz2018nonlinear}%
  \BibitemOpen
  \bibfield  {author} {\bibinfo {author} {\bibfnamefont {S.~H.}\ \bibnamefont
  {Strogatz}},\ }\href@noop {} {\emph {\bibinfo {title} {Nonlinear Dynamics and
  Chaos: with Applications to Physics, Biology, Chemistry, and Engineering}}}\
  (\bibinfo  {publisher} {CRC press},\ \bibinfo {address} {Boca Raton},\
  \bibinfo {year} {2018})\BibitemShut {NoStop}%
\bibitem [{\citenamefont {Hriscu}\ and\ \citenamefont
  {Nazarov}(2013)}]{PhysRevLett.110.097002}%
  \BibitemOpen
  \bibfield  {author} {\bibinfo {author} {\bibfnamefont {A.~M.}\ \bibnamefont
  {Hriscu}}\ and\ \bibinfo {author} {\bibfnamefont {Y.~V.}\ \bibnamefont
  {Nazarov}},\ }\bibfield  {title} {\bibinfo {title} {Quantum synchronization
  of conjugated variables in a superconducting device leads to the fundamental
  resistance quantization},\ }\href
  {https://doi.org/10.1103/PhysRevLett.110.097002} {\bibfield  {journal}
  {\bibinfo  {journal} {Phys. Rev. Lett.}\ }\textbf {\bibinfo {volume} {110}},\
  \bibinfo {pages} {097002} (\bibinfo {year} {2013})}\BibitemShut {NoStop}%
\bibitem [{\citenamefont {Nigg}(2018)}]{PhysRevA.97.013811}%
  \BibitemOpen
  \bibfield  {author} {\bibinfo {author} {\bibfnamefont {S.~E.}\ \bibnamefont
  {Nigg}},\ }\bibfield  {title} {\bibinfo {title} {Observing quantum
  synchronization blockade in circuit quantum electrodynamics},\ }\href
  {https://doi.org/10.1103/PhysRevA.97.013811} {\bibfield  {journal} {\bibinfo
  {journal} {Phys. Rev. A}\ }\textbf {\bibinfo {volume} {97}},\ \bibinfo
  {pages} {013811} (\bibinfo {year} {2018})}\BibitemShut {NoStop}%
\bibitem [{\citenamefont {L\"orch}\ \emph {et~al.}(2014)\citenamefont
  {L\"orch}, \citenamefont {Qian}, \citenamefont {Clerk}, \citenamefont
  {Marquardt},\ and\ \citenamefont {Hammerer}}]{PhysRevX.4.011015}%
  \BibitemOpen
  \bibfield  {author} {\bibinfo {author} {\bibfnamefont {N.}~\bibnamefont
  {L\"orch}}, \bibinfo {author} {\bibfnamefont {J.}~\bibnamefont {Qian}},
  \bibinfo {author} {\bibfnamefont {A.}~\bibnamefont {Clerk}}, \bibinfo
  {author} {\bibfnamefont {F.}~\bibnamefont {Marquardt}},\ and\ \bibinfo
  {author} {\bibfnamefont {K.}~\bibnamefont {Hammerer}},\ }\bibfield  {title}
  {\bibinfo {title} {Laser theory for optomechanics: limit cycles in the
  quantum regime},\ }\href {https://doi.org/10.1103/PhysRevX.4.011015}
  {\bibfield  {journal} {\bibinfo  {journal} {Phys. Rev. X}\ }\textbf {\bibinfo
  {volume} {4}},\ \bibinfo {pages} {011015} (\bibinfo {year}
  {2014})}\BibitemShut {NoStop}%
\bibitem [{\citenamefont {Weiss}\ \emph {et~al.}(2016)\citenamefont {Weiss},
  \citenamefont {Kronwald},\ and\ \citenamefont {Marquardt}}]{Weiss_2016}%
  \BibitemOpen
  \bibfield  {author} {\bibinfo {author} {\bibfnamefont {T.}~\bibnamefont
  {Weiss}}, \bibinfo {author} {\bibfnamefont {A.}~\bibnamefont {Kronwald}},\
  and\ \bibinfo {author} {\bibfnamefont {F.}~\bibnamefont {Marquardt}},\
  }\bibfield  {title} {\bibinfo {title} {Noise-induced transitions in
  optomechanical synchronization},\ }\href
  {https://doi.org/10.1088/1367-2630/18/1/013043} {\bibfield  {journal}
  {\bibinfo  {journal} {New J. Phys.}\ }\textbf {\bibinfo {volume} {18}},\
  \bibinfo {pages} {013043} (\bibinfo {year} {2016})}\BibitemShut {NoStop}%
\bibitem [{\citenamefont {Javaloyes}\ \emph {et~al.}(2008)\citenamefont
  {Javaloyes}, \citenamefont {Perrin},\ and\ \citenamefont
  {Politi}}]{PhysRevE.78.011108}%
  \BibitemOpen
  \bibfield  {author} {\bibinfo {author} {\bibfnamefont {J.}~\bibnamefont
  {Javaloyes}}, \bibinfo {author} {\bibfnamefont {M.}~\bibnamefont {Perrin}},\
  and\ \bibinfo {author} {\bibfnamefont {A.}~\bibnamefont {Politi}},\
  }\bibfield  {title} {\bibinfo {title} {Collective atomic recoil laser as a
  synchronization transition},\ }\href
  {https://doi.org/10.1103/PhysRevE.78.011108} {\bibfield  {journal} {\bibinfo
  {journal} {Phys. Rev. E}\ }\textbf {\bibinfo {volume} {78}},\ \bibinfo
  {pages} {011108} (\bibinfo {year} {2008})}\BibitemShut {NoStop}%
\bibitem [{\citenamefont {Xu}\ and\ \citenamefont
  {Holland}(2015)}]{PhysRevLett.114.103601}%
  \BibitemOpen
  \bibfield  {author} {\bibinfo {author} {\bibfnamefont {M.}~\bibnamefont
  {Xu}}\ and\ \bibinfo {author} {\bibfnamefont {M.~J.}\ \bibnamefont
  {Holland}},\ }\bibfield  {title} {\bibinfo {title} {Conditional ramsey
  spectroscopy with synchronized atoms},\ }\href
  {https://doi.org/10.1103/PhysRevLett.114.103601} {\bibfield  {journal}
  {\bibinfo  {journal} {Phys. Rev. Lett.}\ }\textbf {\bibinfo {volume} {114}},\
  \bibinfo {pages} {103601} (\bibinfo {year} {2015})}\BibitemShut {NoStop}%
\bibitem [{\citenamefont {Feldmann}\ and\ \citenamefont
  {Kosloff}(2004)}]{PhysRevE.70.046110}%
  \BibitemOpen
  \bibfield  {author} {\bibinfo {author} {\bibfnamefont {T.}~\bibnamefont
  {Feldmann}}\ and\ \bibinfo {author} {\bibfnamefont {R.}~\bibnamefont
  {Kosloff}},\ }\bibfield  {title} {\bibinfo {title} {Characteristics of the
  limit cycle of a reciprocating quantum heat engine},\ }\href
  {https://doi.org/10.1103/PhysRevE.70.046110} {\bibfield  {journal} {\bibinfo
  {journal} {Phys. Rev. E}\ }\textbf {\bibinfo {volume} {70}},\ \bibinfo
  {pages} {046110} (\bibinfo {year} {2004})}\BibitemShut {NoStop}%
\bibitem [{\citenamefont {Rezek}\ and\ \citenamefont
  {Kosloff}(2006)}]{Rezek_2006}%
  \BibitemOpen
  \bibfield  {author} {\bibinfo {author} {\bibfnamefont {Y.}~\bibnamefont
  {Rezek}}\ and\ \bibinfo {author} {\bibfnamefont {R.}~\bibnamefont
  {Kosloff}},\ }\bibfield  {title} {\bibinfo {title} {Irreversible performance
  of a quantum harmonic heat engine},\ }\href
  {https://doi.org/10.1088/1367-2630/8/5/083} {\bibfield  {journal} {\bibinfo
  {journal} {New J. Phys.}\ }\textbf {\bibinfo {volume} {8}},\ \bibinfo {pages}
  {83} (\bibinfo {year} {2006})}\BibitemShut {NoStop}%
\bibitem [{\citenamefont {Hardal}\ \emph {et~al.}(2017)\citenamefont {Hardal},
  \citenamefont {Aslan}, \citenamefont {Wilson},\ and\ \citenamefont
  {M\"ustecapl\ifmmode \imath \else \i \fi{}o\ifmmode~\breve{g}\else
  \u{g}\fi{}lu}}]{PhysRevE.96.062120}%
  \BibitemOpen
  \bibfield  {author} {\bibinfo {author} {\bibfnamefont {A.~U.~C.}\
  \bibnamefont {Hardal}}, \bibinfo {author} {\bibfnamefont {N.}~\bibnamefont
  {Aslan}}, \bibinfo {author} {\bibfnamefont {C.~M.}\ \bibnamefont {Wilson}},\
  and\ \bibinfo {author} {\bibfnamefont {O.~E.}\ \bibnamefont
  {M\"ustecapl\ifmmode \imath \else \i \fi{}o\ifmmode~\breve{g}\else
  \u{g}\fi{}lu}},\ }\bibfield  {title} {\bibinfo {title} {Quantum heat engine
  with coupled superconducting resonators},\ }\href
  {https://doi.org/10.1103/PhysRevE.96.062120} {\bibfield  {journal} {\bibinfo
  {journal} {Phys. Rev. E}\ }\textbf {\bibinfo {volume} {96}},\ \bibinfo
  {pages} {062120} (\bibinfo {year} {2017})}\BibitemShut {NoStop}%
\bibitem [{\citenamefont {Jaseem}\ \emph {et~al.}(2020)\citenamefont {Jaseem},
  \citenamefont {Hajdu\ifmmode~\check{s}\else \v{s}\fi{}ek}, \citenamefont
  {Vedral}, \citenamefont {Fazio}, \citenamefont {Kwek},\ and\ \citenamefont
  {Vinjanampathy}}]{PhysRevE.101.020201}%
  \BibitemOpen
  \bibfield  {author} {\bibinfo {author} {\bibfnamefont {N.}~\bibnamefont
  {Jaseem}}, \bibinfo {author} {\bibfnamefont {M.}~\bibnamefont
  {Hajdu\ifmmode~\check{s}\else \v{s}\fi{}ek}}, \bibinfo {author}
  {\bibfnamefont {V.}~\bibnamefont {Vedral}}, \bibinfo {author} {\bibfnamefont
  {R.}~\bibnamefont {Fazio}}, \bibinfo {author} {\bibfnamefont {L.-C.}\
  \bibnamefont {Kwek}},\ and\ \bibinfo {author} {\bibfnamefont
  {S.}~\bibnamefont {Vinjanampathy}},\ }\bibfield  {title} {\bibinfo {title}
  {Quantum synchronization in nanoscale heat engines},\ }\href
  {https://doi.org/10.1103/PhysRevE.101.020201} {\bibfield  {journal} {\bibinfo
   {journal} {Phys. Rev. E}\ }\textbf {\bibinfo {volume} {101}},\ \bibinfo
  {pages} {020201} (\bibinfo {year} {2020})}\BibitemShut {NoStop}%
\bibitem [{\citenamefont {Crescini}\ \emph {et~al.}(2023)\citenamefont
  {Crescini}, \citenamefont {Cailleaux}, \citenamefont {Guichard},
  \citenamefont {Naud}, \citenamefont {Buisson}, \citenamefont {W.~Murch},\
  and\ \citenamefont {Roch}}]{crescini2023evidence}%
  \BibitemOpen
  \bibfield  {author} {\bibinfo {author} {\bibfnamefont {N.}~\bibnamefont
  {Crescini}}, \bibinfo {author} {\bibfnamefont {S.}~\bibnamefont {Cailleaux}},
  \bibinfo {author} {\bibfnamefont {W.}~\bibnamefont {Guichard}}, \bibinfo
  {author} {\bibfnamefont {C.}~\bibnamefont {Naud}}, \bibinfo {author}
  {\bibfnamefont {O.}~\bibnamefont {Buisson}}, \bibinfo {author} {\bibfnamefont
  {K.}~\bibnamefont {W.~Murch}},\ and\ \bibinfo {author} {\bibfnamefont
  {N.}~\bibnamefont {Roch}},\ }\bibfield  {title} {\bibinfo {title} {Evidence
  of dual {Shapiro} steps in a {Josephson} junction array},\ }\href
  {https://doi.org/https://doi.org/10.1038/s41567-023-01961-4} {\bibfield
  {journal} {\bibinfo  {journal} {Nat. Phys.}\ ,\ \bibinfo {pages} {1}}
  (\bibinfo {year} {2023})}\BibitemShut {NoStop}%
\bibitem [{\citenamefont {Ladd}\ \emph {et~al.}(2018)\citenamefont {Ladd},
  \citenamefont {Bachert}, \citenamefont {Meyerspeer}, \citenamefont {Moser},
  \citenamefont {Nagel}, \citenamefont {Norris}, \citenamefont {Schmitter},
  \citenamefont {Speck}, \citenamefont {Straub},\ and\ \citenamefont
  {Zaiss}}]{LADD20181}%
  \BibitemOpen
  \bibfield  {author} {\bibinfo {author} {\bibfnamefont {M.~E.}\ \bibnamefont
  {Ladd}}, \bibinfo {author} {\bibfnamefont {P.}~\bibnamefont {Bachert}},
  \bibinfo {author} {\bibfnamefont {M.}~\bibnamefont {Meyerspeer}}, \bibinfo
  {author} {\bibfnamefont {E.}~\bibnamefont {Moser}}, \bibinfo {author}
  {\bibfnamefont {A.~M.}\ \bibnamefont {Nagel}}, \bibinfo {author}
  {\bibfnamefont {D.~G.}\ \bibnamefont {Norris}}, \bibinfo {author}
  {\bibfnamefont {S.}~\bibnamefont {Schmitter}}, \bibinfo {author}
  {\bibfnamefont {O.}~\bibnamefont {Speck}}, \bibinfo {author} {\bibfnamefont
  {S.}~\bibnamefont {Straub}},\ and\ \bibinfo {author} {\bibfnamefont
  {M.}~\bibnamefont {Zaiss}},\ }\bibfield  {title} {\bibinfo {title} {Pros and
  cons of ultra-high-field {MRI/MRS} for human application},\ }\href
  {https://doi.org/https://doi.org/10.1016/j.pnmrs.2018.06.001} {\bibfield
  {journal} {\bibinfo  {journal} {Prog. Nucl. Magn. Reson. Spectrosc.}\
  }\textbf {\bibinfo {volume} {109}},\ \bibinfo {pages} {1} (\bibinfo {year}
  {2018})}\BibitemShut {NoStop}%
\bibitem [{\citenamefont {Liu}\ \emph {et~al.}(2021)\citenamefont {Liu},
  \citenamefont {Quan}, \citenamefont {Xiang}, \citenamefont {Hong},
  \citenamefont {Cao}, \citenamefont {Liu}, \citenamefont {Dong},\ and\
  \citenamefont {Zhang}}]{doi:10.1063/5.0061478}%
  \BibitemOpen
  \bibfield  {author} {\bibinfo {author} {\bibfnamefont {Y.}~\bibnamefont
  {Liu}}, \bibinfo {author} {\bibfnamefont {R.}~\bibnamefont {Quan}}, \bibinfo
  {author} {\bibfnamefont {X.}~\bibnamefont {Xiang}}, \bibinfo {author}
  {\bibfnamefont {H.}~\bibnamefont {Hong}}, \bibinfo {author} {\bibfnamefont
  {M.}~\bibnamefont {Cao}}, \bibinfo {author} {\bibfnamefont {T.}~\bibnamefont
  {Liu}}, \bibinfo {author} {\bibfnamefont {R.}~\bibnamefont {Dong}},\ and\
  \bibinfo {author} {\bibfnamefont {S.}~\bibnamefont {Zhang}},\ }\bibfield
  {title} {\bibinfo {title} {Quantum clock synchronization over 20-km multiple
  segmented fibers with frequency-correlated photon pairs and hom
  interference},\ }\href {https://doi.org/10.1063/5.0061478} {\bibfield
  {journal} {\bibinfo  {journal} {Appl. Phys. Lett.}\ }\textbf {\bibinfo
  {volume} {119}},\ \bibinfo {pages} {144003} (\bibinfo {year}
  {2021})}\BibitemShut {NoStop}%
\bibitem [{\citenamefont {Schwarzhans}\ \emph {et~al.}(2021)\citenamefont
  {Schwarzhans}, \citenamefont {Lock}, \citenamefont {Erker}, \citenamefont
  {Friis},\ and\ \citenamefont {Huber}}]{PhysRevX.11.011046}%
  \BibitemOpen
  \bibfield  {author} {\bibinfo {author} {\bibfnamefont {E.}~\bibnamefont
  {Schwarzhans}}, \bibinfo {author} {\bibfnamefont {M.~P.~E.}\ \bibnamefont
  {Lock}}, \bibinfo {author} {\bibfnamefont {P.}~\bibnamefont {Erker}},
  \bibinfo {author} {\bibfnamefont {N.}~\bibnamefont {Friis}},\ and\ \bibinfo
  {author} {\bibfnamefont {M.}~\bibnamefont {Huber}},\ }\bibfield  {title}
  {\bibinfo {title} {Autonomous temporal probability concentration: Clockworks
  and the second law of thermodynamics},\ }\href
  {https://doi.org/10.1103/PhysRevX.11.011046} {\bibfield  {journal} {\bibinfo
  {journal} {Phys. Rev. X}\ }\textbf {\bibinfo {volume} {11}},\ \bibinfo
  {pages} {011046} (\bibinfo {year} {2021})}\BibitemShut {NoStop}%
\bibitem [{\citenamefont {Roulet}\ and\ \citenamefont
  {Bruder}(2018)}]{PhysRevLett.121.053601}%
  \BibitemOpen
  \bibfield  {author} {\bibinfo {author} {\bibfnamefont {A.}~\bibnamefont
  {Roulet}}\ and\ \bibinfo {author} {\bibfnamefont {C.}~\bibnamefont
  {Bruder}},\ }\bibfield  {title} {\bibinfo {title} {Synchronizing the smallest
  possible system},\ }\href {https://doi.org/10.1103/PhysRevLett.121.053601}
  {\bibfield  {journal} {\bibinfo  {journal} {Phys. Rev. Lett.}\ }\textbf
  {\bibinfo {volume} {121}},\ \bibinfo {pages} {053601} (\bibinfo {year}
  {2018})}\BibitemShut {NoStop}%
\bibitem [{\citenamefont {Parra-L\'opez}\ and\ \citenamefont
  {Bergli}(2020)}]{PhysRevA.101.062104}%
  \BibitemOpen
  \bibfield  {author} {\bibinfo {author} {\bibfnamefont {A.}~\bibnamefont
  {Parra-L\'opez}}\ and\ \bibinfo {author} {\bibfnamefont {J.}~\bibnamefont
  {Bergli}},\ }\bibfield  {title} {\bibinfo {title} {Synchronization in
  two-level quantum systems},\ }\href
  {https://doi.org/10.1103/PhysRevA.101.062104} {\bibfield  {journal} {\bibinfo
   {journal} {Phys. Rev. A}\ }\textbf {\bibinfo {volume} {101}},\ \bibinfo
  {pages} {062104} (\bibinfo {year} {2020})}\BibitemShut {NoStop}%
\bibitem [{\citenamefont {Lee}\ and\ \citenamefont
  {Sadeghpour}(2013)}]{PhysRevLett.111.234101}%
  \BibitemOpen
  \bibfield  {author} {\bibinfo {author} {\bibfnamefont {T.~E.}\ \bibnamefont
  {Lee}}\ and\ \bibinfo {author} {\bibfnamefont {H.~R.}\ \bibnamefont
  {Sadeghpour}},\ }\bibfield  {title} {\bibinfo {title} {Quantum
  synchronization of quantum van der {Pol} oscillators with trapped ions},\
  }\href {https://doi.org/10.1103/PhysRevLett.111.234101} {\bibfield  {journal}
  {\bibinfo  {journal} {Phys. Rev. Lett.}\ }\textbf {\bibinfo {volume} {111}},\
  \bibinfo {pages} {234101} (\bibinfo {year} {2013})}\BibitemShut {NoStop}%
\bibitem [{\citenamefont {Walter}\ \emph {et~al.}(2014)\citenamefont {Walter},
  \citenamefont {Nunnenkamp},\ and\ \citenamefont
  {Bruder}}]{PhysRevLett.112.094102}%
  \BibitemOpen
  \bibfield  {author} {\bibinfo {author} {\bibfnamefont {S.}~\bibnamefont
  {Walter}}, \bibinfo {author} {\bibfnamefont {A.}~\bibnamefont {Nunnenkamp}},\
  and\ \bibinfo {author} {\bibfnamefont {C.}~\bibnamefont {Bruder}},\
  }\bibfield  {title} {\bibinfo {title} {Quantum synchronization of a driven
  self-sustained oscillator},\ }\href
  {https://doi.org/10.1103/PhysRevLett.112.094102} {\bibfield  {journal}
  {\bibinfo  {journal} {Phys. Rev. Lett.}\ }\textbf {\bibinfo {volume} {112}},\
  \bibinfo {pages} {094102} (\bibinfo {year} {2014})}\BibitemShut {NoStop}%
\bibitem [{\citenamefont {Laskar}\ \emph {et~al.}(2020)\citenamefont {Laskar},
  \citenamefont {Adhikary}, \citenamefont {Mondal}, \citenamefont {Katiyar},
  \citenamefont {Vinjanampathy},\ and\ \citenamefont
  {Ghosh}}]{PhysRevLett.125.013601}%
  \BibitemOpen
  \bibfield  {author} {\bibinfo {author} {\bibfnamefont {A.~W.}\ \bibnamefont
  {Laskar}}, \bibinfo {author} {\bibfnamefont {P.}~\bibnamefont {Adhikary}},
  \bibinfo {author} {\bibfnamefont {S.}~\bibnamefont {Mondal}}, \bibinfo
  {author} {\bibfnamefont {P.}~\bibnamefont {Katiyar}}, \bibinfo {author}
  {\bibfnamefont {S.}~\bibnamefont {Vinjanampathy}},\ and\ \bibinfo {author}
  {\bibfnamefont {S.}~\bibnamefont {Ghosh}},\ }\bibfield  {title} {\bibinfo
  {title} {Observation of quantum phase synchronization in spin-1 atoms},\
  }\href {https://doi.org/10.1103/PhysRevLett.125.013601} {\bibfield  {journal}
  {\bibinfo  {journal} {Phys. Rev. Lett.}\ }\textbf {\bibinfo {volume} {125}},\
  \bibinfo {pages} {013601} (\bibinfo {year} {2020})}\BibitemShut {NoStop}%
\bibitem [{\citenamefont {Kongkhambut}\ \emph {et~al.}(2022)\citenamefont
  {Kongkhambut}, \citenamefont {Skulte}, \citenamefont {Mathey}, \citenamefont
  {Cosme}, \citenamefont {Hemmerich},\ and\ \citenamefont
  {Keßler}}]{doi:10.1126/science.abo3382}%
  \BibitemOpen
  \bibfield  {author} {\bibinfo {author} {\bibfnamefont {P.}~\bibnamefont
  {Kongkhambut}}, \bibinfo {author} {\bibfnamefont {J.}~\bibnamefont {Skulte}},
  \bibinfo {author} {\bibfnamefont {L.}~\bibnamefont {Mathey}}, \bibinfo
  {author} {\bibfnamefont {J.~G.}\ \bibnamefont {Cosme}}, \bibinfo {author}
  {\bibfnamefont {A.}~\bibnamefont {Hemmerich}},\ and\ \bibinfo {author}
  {\bibfnamefont {H.}~\bibnamefont {Keßler}},\ }\bibfield  {title} {\bibinfo
  {title} {Observation of a continuous time crystal},\ }\href
  {https://doi.org/10.1126/science.abo3382} {\bibfield  {journal} {\bibinfo
  {journal} {Science}\ }\textbf {\bibinfo {volume} {377}},\ \bibinfo {pages}
  {670} (\bibinfo {year} {2022})}\BibitemShut {NoStop}%
\bibitem [{\citenamefont {Koppenh\"ofer}\ \emph {et~al.}(2020)\citenamefont
  {Koppenh\"ofer}, \citenamefont {Bruder},\ and\ \citenamefont
  {Roulet}}]{PhysRevResearch.2.023026}%
  \BibitemOpen
  \bibfield  {author} {\bibinfo {author} {\bibfnamefont {M.}~\bibnamefont
  {Koppenh\"ofer}}, \bibinfo {author} {\bibfnamefont {C.}~\bibnamefont
  {Bruder}},\ and\ \bibinfo {author} {\bibfnamefont {A.}~\bibnamefont
  {Roulet}},\ }\bibfield  {title} {\bibinfo {title} {Quantum synchronization on
  the {IBM} {Q} system},\ }\href
  {https://doi.org/10.1103/PhysRevResearch.2.023026} {\bibfield  {journal}
  {\bibinfo  {journal} {Phys. Rev. Res.}\ }\textbf {\bibinfo {volume} {2}},\
  \bibinfo {pages} {023026} (\bibinfo {year} {2020})}\BibitemShut {NoStop}%
\bibitem [{\citenamefont {Setoyama}\ and\ \citenamefont
  {Hasegawa}(2024)}]{PhysRevLett.132.093602}%
  \BibitemOpen
  \bibfield  {author} {\bibinfo {author} {\bibfnamefont {W.}~\bibnamefont
  {Setoyama}}\ and\ \bibinfo {author} {\bibfnamefont {Y.}~\bibnamefont
  {Hasegawa}},\ }\bibfield  {title} {\bibinfo {title} {Lie algebraic quantum
  phase reduction},\ }\href {https://doi.org/10.1103/PhysRevLett.132.093602}
  {\bibfield  {journal} {\bibinfo  {journal} {Phys. Rev. Lett.}\ }\textbf
  {\bibinfo {volume} {132}},\ \bibinfo {pages} {093602} (\bibinfo {year}
  {2024})}\BibitemShut {NoStop}%
\bibitem [{\citenamefont {Gisin}\ and\ \citenamefont
  {Percival}(1992)}]{Gisin_1992}%
  \BibitemOpen
  \bibfield  {author} {\bibinfo {author} {\bibfnamefont {N.}~\bibnamefont
  {Gisin}}\ and\ \bibinfo {author} {\bibfnamefont {I.~C.}\ \bibnamefont
  {Percival}},\ }\bibfield  {title} {\bibinfo {title} {The quantum-state
  diffusion model applied to open systems},\ }\href
  {https://doi.org/10.1088/0305-4470/25/21/023} {\bibfield  {journal} {\bibinfo
   {journal} {J. Phys. A: Math. Gen.}\ }\textbf {\bibinfo {volume} {25}},\
  \bibinfo {pages} {5677} (\bibinfo {year} {1992})}\BibitemShut {NoStop}%
\bibitem [{\citenamefont {Gardiner}\ \emph {et~al.}(2004)\citenamefont
  {Gardiner}, \citenamefont {Zoller},\ and\ \citenamefont
  {Zoller}}]{gardiner2004quantum}%
  \BibitemOpen
  \bibfield  {author} {\bibinfo {author} {\bibfnamefont {C.}~\bibnamefont
  {Gardiner}}, \bibinfo {author} {\bibfnamefont {P.}~\bibnamefont {Zoller}},\
  and\ \bibinfo {author} {\bibfnamefont {P.}~\bibnamefont {Zoller}},\
  }\href@noop {} {\emph {\bibinfo {title} {{Quantum Noise: A Handbook of
  Markovian and Non-Markovian Quantum Stochastic Methods with Applications to
  Quantum Optics}}}}\ (\bibinfo  {publisher} {Springer Science \& Business
  Media},\ \bibinfo {year} {2004})\BibitemShut {NoStop}%
\bibitem [{\citenamefont {Barchielli}\ and\ \citenamefont
  {Gregoratti}(2009)}]{barchielli2009quantum}%
  \BibitemOpen
  \bibfield  {author} {\bibinfo {author} {\bibfnamefont {A.}~\bibnamefont
  {Barchielli}}\ and\ \bibinfo {author} {\bibfnamefont {M.}~\bibnamefont
  {Gregoratti}},\ }\href@noop {} {\emph {\bibinfo {title} {{Quantum
  Trajectories and Measurements in Continuous Time: the Diffusive Case}}}},\
  Vol.\ \bibinfo {volume} {782}\ (\bibinfo  {publisher} {Springer},\ \bibinfo
  {address} {Berlin},\ \bibinfo {year} {2009})\BibitemShut {NoStop}%
\bibitem [{\citenamefont {Kato}\ \emph {et~al.}(2019)\citenamefont {Kato},
  \citenamefont {Yamamoto},\ and\ \citenamefont
  {Nakao}}]{PhysRevResearch.1.033012}%
  \BibitemOpen
  \bibfield  {author} {\bibinfo {author} {\bibfnamefont {Y.}~\bibnamefont
  {Kato}}, \bibinfo {author} {\bibfnamefont {N.}~\bibnamefont {Yamamoto}},\
  and\ \bibinfo {author} {\bibfnamefont {H.}~\bibnamefont {Nakao}},\ }\bibfield
   {title} {\bibinfo {title} {Semiclassical phase reduction theory for quantum
  synchronization},\ }\href {https://doi.org/10.1103/PhysRevResearch.1.033012}
  {\bibfield  {journal} {\bibinfo  {journal} {Phys. Rev. Res.}\ }\textbf
  {\bibinfo {volume} {1}},\ \bibinfo {pages} {033012} (\bibinfo {year}
  {2019})}\BibitemShut {NoStop}%
\bibitem [{\citenamefont {Lindblad}(1976)}]{lindblad1976generators}%
  \BibitemOpen
  \bibfield  {author} {\bibinfo {author} {\bibfnamefont {G.}~\bibnamefont
  {Lindblad}},\ }\bibfield  {title} {\bibinfo {title} {On the generators of
  quantum dynamical semigroups},\ }\href {https://doi.org/10.1007/BF01608499}
  {\bibfield  {journal} {\bibinfo  {journal} {Commun. Math. Phys.}\ }\textbf
  {\bibinfo {volume} {48}},\ \bibinfo {pages} {119} (\bibinfo {year}
  {1976})}\BibitemShut {NoStop}%
\bibitem [{\citenamefont {Breuer}\ \emph {et~al.}(2002)\citenamefont {Breuer},
  \citenamefont {Petruccione},\ and\ \citenamefont
  {Petruccione}}]{breuer2002theory}%
  \BibitemOpen
  \bibfield  {author} {\bibinfo {author} {\bibfnamefont {H.}~\bibnamefont
  {Breuer}}, \bibinfo {author} {\bibfnamefont {F.}~\bibnamefont
  {Petruccione}},\ and\ \bibinfo {author} {\bibfnamefont {S.}~\bibnamefont
  {Petruccione}},\ }\href@noop {} {\emph {\bibinfo {title} {The Theory of Open
  Quantum Systems}}}\ (\bibinfo  {publisher} {Oxford University Press},\
  \bibinfo {address} {New York},\ \bibinfo {year} {2002})\BibitemShut {NoStop}%
\bibitem [{\citenamefont {Carmichael}(2009)}]{carmichael2009open}%
  \BibitemOpen
  \bibfield  {author} {\bibinfo {author} {\bibfnamefont {H.}~\bibnamefont
  {Carmichael}},\ }\href@noop {} {\emph {\bibinfo {title} {An Open Systems
  Approach to Quantum Optics: Lectures Presented at the {Universit{\'e}} Libre
  de {Bruxelles}}}},\ Lecture Notes in Physics Monographs\ (\bibinfo
  {publisher} {Springer},\ \bibinfo {address} {Berlin; New York},\ \bibinfo
  {year} {2009})\BibitemShut {NoStop}%
\bibitem [{\citenamefont {Wiseman}\ and\ \citenamefont
  {Milburn}(2009)}]{wiseman2009quantum}%
  \BibitemOpen
  \bibfield  {author} {\bibinfo {author} {\bibfnamefont {H.~M.}\ \bibnamefont
  {Wiseman}}\ and\ \bibinfo {author} {\bibfnamefont {G.~J.}\ \bibnamefont
  {Milburn}},\ }\href@noop {} {\emph {\bibinfo {title} {Quantum measurement and
  control}}}\ (\bibinfo  {publisher} {Cambridge university press},\ \bibinfo
  {year} {2009})\BibitemShut {NoStop}%
\bibitem [{\citenamefont {Jacobs}\ and\ \citenamefont
  {Steck}(2006)}]{jacobs2006straightforward}%
  \BibitemOpen
  \bibfield  {author} {\bibinfo {author} {\bibfnamefont {K.}~\bibnamefont
  {Jacobs}}\ and\ \bibinfo {author} {\bibfnamefont {D.~A.}\ \bibnamefont
  {Steck}},\ }\bibfield  {title} {\bibinfo {title} {A straightforward
  introduction to continuous quantum measurement},\ }\href
  {https://www.tandfonline.com/doi/full/10.1080/00107510601101934} {\bibfield
  {journal} {\bibinfo  {journal} {Contemp. Phys.}\ }\textbf {\bibinfo {volume}
  {47}},\ \bibinfo {pages} {279} (\bibinfo {year} {2006})}\BibitemShut
  {NoStop}%
\bibitem [{\citenamefont {Guerlin}\ \emph {et~al.}(2007)\citenamefont
  {Guerlin}, \citenamefont {Bernu}, \citenamefont {Deleglise}, \citenamefont
  {Sayrin}, \citenamefont {Gleyzes}, \citenamefont {Kuhr}, \citenamefont
  {Brune}, \citenamefont {Raimond},\ and\ \citenamefont
  {Haroche}}]{guerlin2007progressive}%
  \BibitemOpen
  \bibfield  {author} {\bibinfo {author} {\bibfnamefont {C.}~\bibnamefont
  {Guerlin}}, \bibinfo {author} {\bibfnamefont {J.}~\bibnamefont {Bernu}},
  \bibinfo {author} {\bibfnamefont {S.}~\bibnamefont {Deleglise}}, \bibinfo
  {author} {\bibfnamefont {C.}~\bibnamefont {Sayrin}}, \bibinfo {author}
  {\bibfnamefont {S.}~\bibnamefont {Gleyzes}}, \bibinfo {author} {\bibfnamefont
  {S.}~\bibnamefont {Kuhr}}, \bibinfo {author} {\bibfnamefont {M.}~\bibnamefont
  {Brune}}, \bibinfo {author} {\bibfnamefont {J.-M.}\ \bibnamefont {Raimond}},\
  and\ \bibinfo {author} {\bibfnamefont {S.}~\bibnamefont {Haroche}},\
  }\bibfield  {title} {\bibinfo {title} {Progressive field-state collapse and
  quantum non-demolition photon counting},\ }\href
  {https://www.nature.com/articles/nature06057} {\bibfield  {journal} {\bibinfo
   {journal} {Nature}\ }\textbf {\bibinfo {volume} {448}},\ \bibinfo {pages}
  {889} (\bibinfo {year} {2007})}\BibitemShut {NoStop}%
\bibitem [{\citenamefont {Murch}\ \emph {et~al.}(2013)\citenamefont {Murch},
  \citenamefont {Weber}, \citenamefont {Macklin},\ and\ \citenamefont
  {Siddiqi}}]{murch2013observing}%
  \BibitemOpen
  \bibfield  {author} {\bibinfo {author} {\bibfnamefont {K.}~\bibnamefont
  {Murch}}, \bibinfo {author} {\bibfnamefont {S.}~\bibnamefont {Weber}},
  \bibinfo {author} {\bibfnamefont {C.}~\bibnamefont {Macklin}},\ and\ \bibinfo
  {author} {\bibfnamefont {I.}~\bibnamefont {Siddiqi}},\ }\bibfield  {title}
  {\bibinfo {title} {Observing single quantum trajectories of a superconducting
  quantum bit},\ }\href {https://www.nature.com/articles/nature12539}
  {\bibfield  {journal} {\bibinfo  {journal} {Nature}\ }\textbf {\bibinfo
  {volume} {502}},\ \bibinfo {pages} {211} (\bibinfo {year}
  {2013})}\BibitemShut {NoStop}%
\bibitem [{\citenamefont {Rossi}\ \emph {et~al.}(2019)\citenamefont {Rossi},
  \citenamefont {Mason}, \citenamefont {Chen},\ and\ \citenamefont
  {Schliesser}}]{PhysRevLett.123.163601}%
  \BibitemOpen
  \bibfield  {author} {\bibinfo {author} {\bibfnamefont {M.}~\bibnamefont
  {Rossi}}, \bibinfo {author} {\bibfnamefont {D.}~\bibnamefont {Mason}},
  \bibinfo {author} {\bibfnamefont {J.}~\bibnamefont {Chen}},\ and\ \bibinfo
  {author} {\bibfnamefont {A.}~\bibnamefont {Schliesser}},\ }\bibfield  {title}
  {\bibinfo {title} {Observing and verifying the quantum trajectory of a
  mechanical resonator},\ }\href
  {https://doi.org/10.1103/PhysRevLett.123.163601} {\bibfield  {journal}
  {\bibinfo  {journal} {Phys. Rev. Lett.}\ }\textbf {\bibinfo {volume} {123}},\
  \bibinfo {pages} {163601} (\bibinfo {year} {2019})}\BibitemShut {NoStop}%
\bibitem [{\citenamefont {Teramae}\ \emph {et~al.}(2009)\citenamefont
  {Teramae}, \citenamefont {Nakao},\ and\ \citenamefont
  {Ermentrout}}]{PhysRevLett.102.194102}%
  \BibitemOpen
  \bibfield  {author} {\bibinfo {author} {\bibfnamefont {J.~N.}\ \bibnamefont
  {Teramae}}, \bibinfo {author} {\bibfnamefont {H.}~\bibnamefont {Nakao}},\
  and\ \bibinfo {author} {\bibfnamefont {G.~B.}\ \bibnamefont {Ermentrout}},\
  }\bibfield  {title} {\bibinfo {title} {Stochastic phase reduction for a
  general class of noisy limit cycle oscillators},\ }\href
  {https://doi.org/10.1103/PhysRevLett.102.194102} {\bibfield  {journal}
  {\bibinfo  {journal} {Phys. Rev. Lett.}\ }\textbf {\bibinfo {volume} {102}},\
  \bibinfo {pages} {194102} (\bibinfo {year} {2009})}\BibitemShut {NoStop}%
\bibitem [{\citenamefont {Nakao}\ \emph {et~al.}(2007)\citenamefont {Nakao},
  \citenamefont {Arai},\ and\ \citenamefont
  {Kawamura}}]{PhysRevLett.98.184101}%
  \BibitemOpen
  \bibfield  {author} {\bibinfo {author} {\bibfnamefont {H.}~\bibnamefont
  {Nakao}}, \bibinfo {author} {\bibfnamefont {K.}~\bibnamefont {Arai}},\ and\
  \bibinfo {author} {\bibfnamefont {Y.}~\bibnamefont {Kawamura}},\ }\bibfield
  {title} {\bibinfo {title} {Noise-induced synchronization and clustering in
  ensembles of uncoupled limit-cycle oscillators},\ }\href
  {https://doi.org/10.1103/PhysRevLett.98.184101} {\bibfield  {journal}
  {\bibinfo  {journal} {Phys. Rev. Lett.}\ }\textbf {\bibinfo {volume} {98}},\
  \bibinfo {pages} {184101} (\bibinfo {year} {2007})}\BibitemShut {NoStop}%
\bibitem [{\citenamefont {Yoshimura}\ and\ \citenamefont
  {Arai}(2008)}]{PhysRevLett.101.154101}%
  \BibitemOpen
  \bibfield  {author} {\bibinfo {author} {\bibfnamefont {K.}~\bibnamefont
  {Yoshimura}}\ and\ \bibinfo {author} {\bibfnamefont {K.}~\bibnamefont
  {Arai}},\ }\bibfield  {title} {\bibinfo {title} {Phase reduction of
  stochastic limit cycle oscillators},\ }\href
  {https://doi.org/10.1103/PhysRevLett.101.154101} {\bibfield  {journal}
  {\bibinfo  {journal} {Phys. Rev. Lett.}\ }\textbf {\bibinfo {volume} {101}},\
  \bibinfo {pages} {154101} (\bibinfo {year} {2008})}\BibitemShut {NoStop}%
\bibitem [{\citenamefont {Goldobin}\ and\ \citenamefont
  {Pikovsky}(2005)}]{GOLDOBIN2005126}%
  \BibitemOpen
  \bibfield  {author} {\bibinfo {author} {\bibfnamefont {D.}~\bibnamefont
  {Goldobin}}\ and\ \bibinfo {author} {\bibfnamefont {A.}~\bibnamefont
  {Pikovsky}},\ }\bibfield  {title} {\bibinfo {title} {Synchronization of
  self-sustained oscillators by common white noise},\ }\href
  {https://doi.org/https://doi.org/10.1016/j.physa.2004.12.014} {\bibfield
  {journal} {\bibinfo  {journal} {Physica A}\ }\textbf {\bibinfo {volume}
  {351}},\ \bibinfo {pages} {126} (\bibinfo {year} {2005})}\BibitemShut
  {NoStop}%
\bibitem [{\citenamefont {Nakao}(2016)}]{doi:10.1080/00107514.2015.1094987}%
  \BibitemOpen
  \bibfield  {author} {\bibinfo {author} {\bibfnamefont {H.}~\bibnamefont
  {Nakao}},\ }\bibfield  {title} {\bibinfo {title} {Phase reduction approach to
  synchronisation of nonlinear oscillators},\ }\href
  {https://doi.org/10.1080/00107514.2015.1094987} {\bibfield  {journal}
  {\bibinfo  {journal} {Contemporary Physics}\ }\textbf {\bibinfo {volume}
  {57}},\ \bibinfo {pages} {188} (\bibinfo {year} {2016})}\BibitemShut
  {NoStop}%
\bibitem [{\citenamefont {Neiman}\ and\ \citenamefont
  {Russell}(2002)}]{PhysRevLett.88.138103}%
  \BibitemOpen
  \bibfield  {author} {\bibinfo {author} {\bibfnamefont {A.~B.}\ \bibnamefont
  {Neiman}}\ and\ \bibinfo {author} {\bibfnamefont {D.~F.}\ \bibnamefont
  {Russell}},\ }\bibfield  {title} {\bibinfo {title} {Synchronization of
  noise-induced bursts in noncoupled sensory neurons},\ }\href
  {https://doi.org/10.1103/PhysRevLett.88.138103} {\bibfield  {journal}
  {\bibinfo  {journal} {Phys. Rev. Lett.}\ }\textbf {\bibinfo {volume} {88}},\
  \bibinfo {pages} {138103} (\bibinfo {year} {2002})}\BibitemShut {NoStop}%
\bibitem [{\citenamefont {Zhou}\ \emph {et~al.}(2003)\citenamefont {Zhou},
  \citenamefont {Kurths}, \citenamefont {Allaria}, \citenamefont {Boccaletti},
  \citenamefont {Meucci},\ and\ \citenamefont {Arecchi}}]{PhysRevE.67.066220}%
  \BibitemOpen
  \bibfield  {author} {\bibinfo {author} {\bibfnamefont {C.~S.}\ \bibnamefont
  {Zhou}}, \bibinfo {author} {\bibfnamefont {J.}~\bibnamefont {Kurths}},
  \bibinfo {author} {\bibfnamefont {E.}~\bibnamefont {Allaria}}, \bibinfo
  {author} {\bibfnamefont {S.}~\bibnamefont {Boccaletti}}, \bibinfo {author}
  {\bibfnamefont {R.}~\bibnamefont {Meucci}},\ and\ \bibinfo {author}
  {\bibfnamefont {F.~T.}\ \bibnamefont {Arecchi}},\ }\bibfield  {title}
  {\bibinfo {title} {Constructive effects of noise in homoclinic chaotic
  systems},\ }\href {https://doi.org/10.1103/PhysRevE.67.066220} {\bibfield
  {journal} {\bibinfo  {journal} {Phys. Rev. E}\ }\textbf {\bibinfo {volume}
  {67}},\ \bibinfo {pages} {066220} (\bibinfo {year} {2003})}\BibitemShut
  {NoStop}%
\bibitem [{\citenamefont {Zhou}\ and\ \citenamefont
  {Kurths}(2003)}]{doi:10.1063/1.1493096}%
  \BibitemOpen
  \bibfield  {author} {\bibinfo {author} {\bibfnamefont {C.}~\bibnamefont
  {Zhou}}\ and\ \bibinfo {author} {\bibfnamefont {J.}~\bibnamefont {Kurths}},\
  }\bibfield  {title} {\bibinfo {title} {Noise-induced synchronization and
  coherence resonance of a {Hodgkin–Huxley} model of thermally sensitive
  neurons},\ }\href {https://doi.org/10.1063/1.1493096} {\bibfield  {journal}
  {\bibinfo  {journal} {Chaos}\ }\textbf {\bibinfo {volume} {13}},\ \bibinfo
  {pages} {401} (\bibinfo {year} {2003})}\BibitemShut {NoStop}%
\bibitem [{\citenamefont {Schmolke}\ and\ \citenamefont
  {Lutz}(2022)}]{PhysRevLett.129.250601}%
  \BibitemOpen
  \bibfield  {author} {\bibinfo {author} {\bibfnamefont {F.}~\bibnamefont
  {Schmolke}}\ and\ \bibinfo {author} {\bibfnamefont {E.}~\bibnamefont
  {Lutz}},\ }\bibfield  {title} {\bibinfo {title} {Noise-induced quantum
  synchronization},\ }\href {https://doi.org/10.1103/PhysRevLett.129.250601}
  {\bibfield  {journal} {\bibinfo  {journal} {Phys. Rev. Lett.}\ }\textbf
  {\bibinfo {volume} {129}},\ \bibinfo {pages} {250601} (\bibinfo {year}
  {2022})}\BibitemShut {NoStop}%
\bibitem [{\citenamefont {Maier}\ \emph {et~al.}(2019)\citenamefont {Maier},
  \citenamefont {Brydges}, \citenamefont {Jurcevic}, \citenamefont {Trautmann},
  \citenamefont {Hempel}, \citenamefont {Lanyon}, \citenamefont {Hauke},
  \citenamefont {Blatt},\ and\ \citenamefont {Roos}}]{PhysRevLett.122.050501}%
  \BibitemOpen
  \bibfield  {author} {\bibinfo {author} {\bibfnamefont {C.}~\bibnamefont
  {Maier}}, \bibinfo {author} {\bibfnamefont {T.}~\bibnamefont {Brydges}},
  \bibinfo {author} {\bibfnamefont {P.}~\bibnamefont {Jurcevic}}, \bibinfo
  {author} {\bibfnamefont {N.}~\bibnamefont {Trautmann}}, \bibinfo {author}
  {\bibfnamefont {C.}~\bibnamefont {Hempel}}, \bibinfo {author} {\bibfnamefont
  {B.~P.}\ \bibnamefont {Lanyon}}, \bibinfo {author} {\bibfnamefont
  {P.}~\bibnamefont {Hauke}}, \bibinfo {author} {\bibfnamefont
  {R.}~\bibnamefont {Blatt}},\ and\ \bibinfo {author} {\bibfnamefont {C.~F.}\
  \bibnamefont {Roos}},\ }\bibfield  {title} {\bibinfo {title}
  {Environment-assisted quantum transport in a 10-qubit network},\ }\href
  {https://doi.org/10.1103/PhysRevLett.122.050501} {\bibfield  {journal}
  {\bibinfo  {journal} {Phys. Rev. Lett.}\ }\textbf {\bibinfo {volume} {122}},\
  \bibinfo {pages} {050501} (\bibinfo {year} {2019})}\BibitemShut {NoStop}%
\bibitem [{\citenamefont {Karimi}\ and\ \citenamefont
  {Pekola}(2017)}]{PhysRevB.96.115408}%
  \BibitemOpen
  \bibfield  {author} {\bibinfo {author} {\bibfnamefont {B.}~\bibnamefont
  {Karimi}}\ and\ \bibinfo {author} {\bibfnamefont {J.~P.}\ \bibnamefont
  {Pekola}},\ }\bibfield  {title} {\bibinfo {title} {Correlated versus
  uncorrelated noise acting on a quantum refrigerator},\ }\href
  {https://doi.org/10.1103/PhysRevB.96.115408} {\bibfield  {journal} {\bibinfo
  {journal} {Phys. Rev. B}\ }\textbf {\bibinfo {volume} {96}},\ \bibinfo
  {pages} {115408} (\bibinfo {year} {2017})}\BibitemShut {NoStop}%
\bibitem [{\citenamefont {Risken}(1996)}]{risken1996fokker}%
  \BibitemOpen
  \bibfield  {author} {\bibinfo {author} {\bibfnamefont {H.}~\bibnamefont
  {Risken}},\ }\bibfield  {title} {\bibinfo {title} {{Fokker-Planck}
  equation},\ }in\ \href@noop {} {\emph {\bibinfo {booktitle} {The
  Fokker-Planck Equation}}}\ (\bibinfo  {publisher} {Springer},\ \bibinfo
  {year} {1996})\ pp.\ \bibinfo {pages} {63--95}\BibitemShut {NoStop}%
\bibitem [{\citenamefont {L\"orch}\ \emph {et~al.}(2016)\citenamefont
  {L\"orch}, \citenamefont {Amitai}, \citenamefont {Nunnenkamp},\ and\
  \citenamefont {Bruder}}]{PhysRevLett.117.073601}%
  \BibitemOpen
  \bibfield  {author} {\bibinfo {author} {\bibfnamefont {N.}~\bibnamefont
  {L\"orch}}, \bibinfo {author} {\bibfnamefont {E.}~\bibnamefont {Amitai}},
  \bibinfo {author} {\bibfnamefont {A.}~\bibnamefont {Nunnenkamp}},\ and\
  \bibinfo {author} {\bibfnamefont {C.}~\bibnamefont {Bruder}},\ }\bibfield
  {title} {\bibinfo {title} {Genuine quantum signatures in synchronization of
  anharmonic self-oscillators},\ }\href
  {https://doi.org/10.1103/PhysRevLett.117.073601} {\bibfield  {journal}
  {\bibinfo  {journal} {Phys. Rev. Lett.}\ }\textbf {\bibinfo {volume} {117}},\
  \bibinfo {pages} {073601} (\bibinfo {year} {2016})}\BibitemShut {NoStop}%
\bibitem [{\citenamefont {L\"orch}\ \emph {et~al.}(2017)\citenamefont
  {L\"orch}, \citenamefont {Nigg}, \citenamefont {Nunnenkamp}, \citenamefont
  {Tiwari},\ and\ \citenamefont {Bruder}}]{PhysRevLett.118.243602}%
  \BibitemOpen
  \bibfield  {author} {\bibinfo {author} {\bibfnamefont {N.}~\bibnamefont
  {L\"orch}}, \bibinfo {author} {\bibfnamefont {S.~E.}\ \bibnamefont {Nigg}},
  \bibinfo {author} {\bibfnamefont {A.}~\bibnamefont {Nunnenkamp}}, \bibinfo
  {author} {\bibfnamefont {R.~P.}\ \bibnamefont {Tiwari}},\ and\ \bibinfo
  {author} {\bibfnamefont {C.}~\bibnamefont {Bruder}},\ }\bibfield  {title}
  {\bibinfo {title} {Quantum synchronization blockade: Energy quantization
  hinders synchronization of identical oscillators},\ }\href
  {https://doi.org/10.1103/PhysRevLett.118.243602} {\bibfield  {journal}
  {\bibinfo  {journal} {Phys. Rev. Lett.}\ }\textbf {\bibinfo {volume} {118}},\
  \bibinfo {pages} {243602} (\bibinfo {year} {2017})}\BibitemShut {NoStop}%
\bibitem [{\citenamefont {Solanki}\ \emph {et~al.}(2023)\citenamefont
  {Solanki}, \citenamefont {Mehdi}, \citenamefont {Hajdu\ifmmode~\check{s}\else
  \v{s}\fi{}ek},\ and\ \citenamefont {Vinjanampathy}}]{PhysRevA.108.022216}%
  \BibitemOpen
  \bibfield  {author} {\bibinfo {author} {\bibfnamefont {P.}~\bibnamefont
  {Solanki}}, \bibinfo {author} {\bibfnamefont {F.~M.}\ \bibnamefont {Mehdi}},
  \bibinfo {author} {\bibfnamefont {M.}~\bibnamefont
  {Hajdu\ifmmode~\check{s}\else \v{s}\fi{}ek}},\ and\ \bibinfo {author}
  {\bibfnamefont {S.}~\bibnamefont {Vinjanampathy}},\ }\bibfield  {title}
  {\bibinfo {title} {Symmetries and synchronization blockade},\ }\href
  {https://doi.org/10.1103/PhysRevA.108.022216} {\bibfield  {journal} {\bibinfo
   {journal} {Phys. Rev. A}\ }\textbf {\bibinfo {volume} {108}},\ \bibinfo
  {pages} {022216} (\bibinfo {year} {2023})}\BibitemShut {NoStop}%
\bibitem [{\citenamefont {Kurebayashi}\ \emph {et~al.}(2013)\citenamefont
  {Kurebayashi}, \citenamefont {Shirasaka},\ and\ \citenamefont
  {Nakao}}]{PhysRevLett.111.214101}%
  \BibitemOpen
  \bibfield  {author} {\bibinfo {author} {\bibfnamefont {W.}~\bibnamefont
  {Kurebayashi}}, \bibinfo {author} {\bibfnamefont {S.}~\bibnamefont
  {Shirasaka}},\ and\ \bibinfo {author} {\bibfnamefont {H.}~\bibnamefont
  {Nakao}},\ }\bibfield  {title} {\bibinfo {title} {Phase reduction method for
  strongly perturbed limit cycle oscillators},\ }\href
  {https://doi.org/10.1103/PhysRevLett.111.214101} {\bibfield  {journal}
  {\bibinfo  {journal} {Phys. Rev. Lett.}\ }\textbf {\bibinfo {volume} {111}},\
  \bibinfo {pages} {214101} (\bibinfo {year} {2013})}\BibitemShut {NoStop}%
\bibitem [{\citenamefont {Wilson}(2020)}]{PhysRevE.101.022220}%
  \BibitemOpen
  \bibfield  {author} {\bibinfo {author} {\bibfnamefont {D.}~\bibnamefont
  {Wilson}},\ }\bibfield  {title} {\bibinfo {title} {Phase-amplitude reduction
  far beyond the weakly perturbed paradigm},\ }\href
  {https://doi.org/10.1103/PhysRevE.101.022220} {\bibfield  {journal} {\bibinfo
   {journal} {Phys. Rev. E}\ }\textbf {\bibinfo {volume} {101}},\ \bibinfo
  {pages} {022220} (\bibinfo {year} {2020})}\BibitemShut {NoStop}%
\end{thebibliography}
%

\end{document}